\documentclass[twocolumn]{aastex631}

\usepackage{soul}
\usepackage{multirow}
\usepackage{comment}
\usepackage{ulem}

\shorttitle{Study of ethanolamine ices}
\shortauthors{Ramachandran et al.}

\graphicspath{{./}{figures/}}

\begin{document}

\title{Experimental and computational study of ethanolamine ices at astrochemical conditions}

\correspondingauthor{R Ramachandran, A Das, B Sivaraman}
\email{ragav.kasak@gmail.com, ankan.das@gmail.com, bhala@prl.res.in}

\author[0000-0002-4122-9501]{R Ramachandran}
\affiliation{Physical Research Laboratory, India.}
\affiliation{Institute of Astronomy Space and Earth Sciences, P 177, CIT Road, Scheme 7m, Kolkata 700054, West Bengal, India}
\author[0000-0001-5720-6294]{Milan Sil}
\affiliation{Univ. Grenoble Alpes, CNRS, IPAG, 38000 Grenoble, France}
\affiliation{Univ Rennes, CNRS, IPR (Institut de Physique de Rennes) - UMR 6251, F-35000 Rennes, France}
\affiliation{Institute of Astronomy Space and Earth Sciences, P 177, CIT Road, Scheme 7m, Kolkata 700054, West Bengal, India}
\author[0000-0003-1602-6849]{Prasanta Gorai}
\affiliation{Rosseland Centre for Solar Physics, University of Oslo, PO Box 1029 Blindern, 0315 Oslo, Norway}
\affiliation{Institute of Theoretical Astrophysics, University of Oslo, PO Box 1029 Blindern, 0315 Oslo, Norway}
\affiliation{Institute of Astronomy Space and Earth Sciences, P 177, CIT Road, Scheme 7m, Kolkata 700054, West Bengal, India}
\author[0000-0003-0022-4226]{J K Meka}
\affiliation{Physical Research Laboratory, India.}
\author[0000-0001-8809-7633]{S Pavithraa}
\affiliation{Department of Applied Chemistry and Institute of Molecular Sciences, National Chiao Tung University, Hsinchu, Taiwan.}
\author[0000-0002-0153-1423]{J -I Lo}
\affiliation{Department of Medical Research, Hualien Tzu Chi Hospital, Buddhist Tzu Chi Medical Foundation, Hualien, Taiwan.}
\author[0000-0002-6528-3721]{S -L Chou}
\affiliation{National Synchrotron Radiation Research Centre, Hsinchu, Taiwan.}
\author[0000-0001-9261-0655]{Y -J Wu}
\affiliation{National Synchrotron Radiation Research Centre, Hsinchu, Taiwan.}
\author[0000-0003-2504-2576]{P Janardhan}
\affiliation{Physical Research Laboratory, India.}
\author[0000-0002-8540-6274]{B -M Cheng}
\affiliation{Department of Medical Research, Hualien Tzu Chi Hospital, Buddhist Tzu Chi Medical Foundation, Hualien, Taiwan.}
\author[0000-0003-1693-453X]{Anil Bhardwaj}
\affiliation{Physical Research Laboratory, India.}
\author[0000-0002-2887-5859]{V\'{i}ctor M. Rivilla}
\affiliation{Centro de Astrobiología (CSIC-INTA), Ctra. de Ajalvir Km. 4, Torrejón de Ardoz, 28850 Madrid, Spain
}
\author[0000-0002-4468-8324]{N J Mason}
\affiliation{Centre for Astrophysics and Planetary Science, School of Physical Sciences,
University of Kent, Canterbury, CT2 7NH, UK.}
\author[0000-0002-2833-0357]{B Sivaraman}$^{\thanks{E-mail: bhala@prl.res.in}}$
\affiliation{Physical Research Laboratory, India.}
\author[0000-0003-4615-602X]{Ankan Das}$^{\thanks{E-mail: ankan.das@gmail.com}}$
\affiliation{Institute of Astronomy Space and Earth Sciences, P 177, CIT Road, Scheme 7m, Kolkata 700054, West Bengal, India}
\affiliation{Max-Planck-Institute for extraterrestrial Physics, P.O. Box 1312 85741 Garching, Germany}

\begin{abstract}
Ethanolamine ($\rm{NH_2CH_2CH_2OH}$) has recently been identified in the molecular cloud G+0.693-0.027, situated in the SgrB2 complex in the Galactic center. However, its presence in other regions, and in particular in star-forming sites, is still elusive. Given its likely role as a precursor to simple amino acids, understanding its presence in the star-forming region is required.
Here, we present the experimentally obtained temperature-dependent spectral features and morphological behavior of pure ethanolamine ices under astrochemical conditions in the $2-12$~\textmu m (MIR) and $120-230$~nm (VUV) regions for the first time. These features would help in understanding its photochemical behavior.
In addition, we present the first chemical models specifically dedicated to ethanolamine. These models include all the discussed chemical routes from the literature, along with the estimated binding energies and activation energies from quantum chemical calculations reported in this work.
We have found that surface reactions: $\rm{CH_2OH+ NH_2CH_2 \rightarrow NH_2CH_2CH_2OH}$ and $\rm{NH_2 + C_2H_4OH \rightarrow NH_2CH_2CH_2OH}$ in warmer regions ($60-90$~K) could play a significant role in the formation of ethanolamine. Our modeled abundance of ethanolamine complements the upper limit of ethanolamine column density estimated in earlier observations in hot core/corino regions. Furthermore, we provide a theoretical estimation of the rotational and distortional constants for various species (such as HNCCO, NH$_2$CHCO, and NH$_2$CH$_2$CO) related to ethanolamine that have not been studied in existing literature. This study could be valuable for identifying these species in the future.
\end{abstract}

%% PLEASE PUT RELEVANT KEYWORDS FROM https://journals.aas.org/keywords-2013/


\section{Introduction} \label{sec:intro}
Nitrogen and oxygen-bearing complex organic molecules (COMs) in the interstellar medium (ISM) are of substantial significance as they are considered the precursors to biomolecules such as amino acids and proteins. Ethanolamine ($\rm{NH_2CH_2CH_2OH}$), also called 2-aminoethanol, glycinol or monoethanolamine, is a probable precursor of simple amino acids like glycine and alanine \citep{Widicus2003,Wirstrom2007}.
Ethanolamine is also a phospholipid precursor, which could play a crucial role in the evolution of the first cellular membranes needed for the emergence of life on early Earth \citep{Rivilla2021,fior22}.
Though several amino acids, including
glycine and alanine, have been discovered on meteorites \citep{Glavin2010,Kvenvolden1971} and comet \citep{altw16}, they have been elusive in the ISM \citep{kuan03,jone07,cunn07,bell08}.
Amino acids in the interstellar medium (ISM) could be formed from protonated amino alcohols and formic acid under suitable conditions \citep{Ehrenfreund2001}. However, a theoretical calculation by \cite{redo15} revealed that the activation barrier of this reaction presents a significant challenge under interstellar conditions.

\cite{Rivilla2021} analyzed molecular data from high-sensitivity, unbiased spectral surveys conducted by the Yebes 40m and the IRAM 30m telescopes. They successfully detected ethanolamine for the first time in the interstellar medium. This detection was made towards the molecular cloud G+0.693–0.027, located in the SgrB2 complex in the Galactic Center. Previously, ethanolamine had been searched for in various star-forming regions, including several hot cores \citep{Widicus2003,Wirstrom2007} and the prototypical hot corino IRAS 16293-2422 B \citep{naza24}. Telescopes such as the Caltech Submillimeter Observatory, Owens Valley Radio Observatory, and Onsala 20m were utilized in these searches. Despite extensive efforts, ethanolamine remained undetected until it was identified by \cite{Rivilla2021}. Before this identification, only upper bounds for its column density had been determined. Furthermore, \cite{Rivilla2021} formulated potential ice-phase formation pathways for ethanolamine based on the previous works of \citep{Wakelam2015,Kameneva2017,Suzuki2018,Krasnokutski2021,ruau16}. However, the formation pathways of interstellar ethanolamine are still poorly known.

Several ice processing experiments \citep{Bernstein2002, Nuevo2008, Elsila2007, Ehrenfreund2001} designed to understand the amino acid formation from simple interstellar molecules have found several organic molecules, including ethanolamine, in the resulting residue (after hydrolysis). 
%\citet{Danger2011} reported the formation of acetonitrile by the VUV irradiation experiment of ethylamine at 20 K through a two-step dehydrogenation process. Under the VUV irradiation, this acetonitrile could react with ammonia in the ice phase, forming amino acetonitrile. 
%\cite{Duvernay2010} conducted Fourier transform infrared (FTIR) spectroscopy and mass spectrometry to study the formation of alpha-aminoethanol, an isomer of ethanolamine. They also reported on its photochemical behavior during the VUV irradiation experiment. However, its interstellar detection remains unfeasible due to the lack of information related to its rotational spectroscopy.

%A recent discovery of ethanolamine in the ISM \citep{Rivilla2021} has ushered the study of the molecule anew.

 %\st{\cite{Rivilla2021} obtained a molecular column density of $(1.51\pm0.07)\times10^{13}$~cm$^{-2}$ for ethanolamine in the G+0.693-0.027 molecular cloud located in the SgrB2 complex in the Galactic Center. They also proposed some ice-phase formation routes of ethanolamine based on some proposed routes by \cite{Wakelam2015, Kameneva2017, Suzuki2018, Krasnokutski2021, ruau16}. }

%Since there are no available rotational spectroscopic data for HNCCO and other precursors, they could not be identified in the region.

\par 
%\st{The laboratory data for ethanolamine and its formation pathways to support the qualitative and quantitative analysis of observational data to confirm the presence of this molecule (and amino alcohols in general) remains incomplete.}
% The advent of the  James Webb Space Telescope (JWST) is allowing us to identify for the first time the IR features of complex molecules in the interstellar icy mantle \citep[and references therein]{roch22}. Moreover, VUV spectroscopy can also be used to identify molecules. Recently \cite{elowitz2021} reported hydrazine on Saturn’s moon Rhea using VUV spectroscopy of astrochemical ices.

Here, we conducted a comprehensive theoretical and experimental study to understand the formation of ethanolamine and its photochemical behavior.
Our paper presents the first experimental results of the IR and VUV study of pure ethanolamine ices under astrophysical conditions. Furthermore, we have developed a chemical model to study the fate of ethanolamine formation 
in hot core/corino regions.

It is organized as follows: Section~\ref{sec:method} discusses the implemented method for this study, Results and discussion are presented in Section~\ref{sec:result}, and the conclusions are summarized in Section~\ref{sec:conclusion}.

\section{Methodology and Computational Details} \label{sec:method}
UV photons (including VUV) are ubiquitous in the ISM and play a vital role in its photochemistry \citep{10.1093/mnras/staa334} and astronomical observations. Infrared spectroscopy is a valuable tool because the vibrational and rotational spectra of molecules could provide unique fingerprints that can be used to identify and characterize them. 
In interpreting interstellar ice's chemical composition, infrared spectra are invaluable. For example, the data gathered by the JWST is often analyzed by considering the comprehensive resources available in the Leiden Ice Database for Astrochemistry \citep[LIDA]{roch22}. The VUV photoabsorption of astrochemical ices helps identify molecules by comparing data from spectra returned by spacecraft in the VUV region. For example, VUV spectroscopy of astrochemical ices facilitated the identification of hydrazine on Saturn’s moon Rhea \citep{elowitz2021}.
%Understanding ethanolamine ice's VUV and IR properties could facilitate its interstellar identification. 
In addition to these experimental studies, a coupled computational approach is needed to comprehend its formation and assess its abundance in interstellar circumstances. This section discusses the experimental and computational methods implemented in this work to study the fate of ethanolamine in interstellar conditions.

\subsection{Experimental studies}
\subsubsection{VUV experiments}
VUV photoabsorption experiments of ethanolamine ices were carried out with the VUV light from beamline BL-03 of the Taiwan Light Source (TLS) at the National Synchrotron Radiation Research Centre (NSRRC), Taiwan. The VUV light from the synchrotron was incident normal to the sample, and the light hence transmitted was passed through a glass window coated with sodium salicylate to convert the incident UV light to visible light and detect it using a photomultiplier tube (Hamamatsu R943-02), operating in a photon-counting mode. Detailed discussions of the experimental setup, liquid sample preparation, and spectral acquisition have been presented in earlier literature \citep{Lu2008}. The ethanolamine molecules of high purity ($>$99$\%$ purity, Sigma Aldrich) were deposited on the LiF substrate at 10 K.
The LiF substrate is chosen because it is UV transparent, especially in the regions of our interest. Upon deposition, a spectrum was recorded at 10 K. The sample was then heated to higher temperatures. 
 %\st{, and spectra were recorded at every 20 K step to observe variations in the VUV spectral signatures between 120 and 230~nm.} 
 After multiple experimental runs, we understand how the spectra change with temperature. Hence, we choose a convenient step size that best captures these changes.  
 A heating rate of 5 K/min is utilized, and upon stabilization at each 20 K interval, the spectrum is captured to analyze variations in VUV spectral signatures ranging from 120 to 230 nm. For all the experiments in VUV, the thickness used was approx $\sim$ 20 ML.

\subsubsection{IR experiments}
The temperature-dependent infrared spectroscopy of the ethanolamine ice was carried out using the Simulator for Astro-molecules at Low Temperature (SALT) facility housed in the Physical Research Laboratory (PRL), Ahmedabad. The setup and other details are mentioned in \citet{ramachandran2023}.
\par Ethanolamine ($>$99$\%$ purity, Sigma Aldrich), available as a liquid, was used. The sample was processed by several freeze-pump-thaw cycles. Due to its good vapor pressure at room temperature (0.54 mbar at 25 $^{\circ}$C), the usual deposition procedure \citep{ramachandran2023,elowitz2021} was followed. In the first step of the experiment, the all-metal leak valve was opened in a controlled way to deposit the ethanolamine molecules on the ZnSe window, which was precooled %\st{cooled} 
to 7~K and subsequently monitored in-situ using FTIR spectroscopy in the mid-IR range ($4000-650$~cm$^{-1}$). The ice was warmed then to higher temperatures with a ramp rate of 5~K~min$^{-1}$, and the spectra were recorded at regular intervals of $5–10$~K until its sublimation. Similar deposition procedures were followed for the deposition at different temperatures (see Table~\ref{tab:expt}). For variable thicknesses, the deposition time was varied while monitoring the absorbance of the most prominent peak of the ice in the mid-IR region. In basic infrared experiments, an ice thickness of approximately 57 ML was employed.

\subsection{Computational studies}
\subsubsection{Binding energies}
One of the main challenges in accurately modeling interstellar chemistry is the limited knowledge of the binding energy (BE) of interstellar species. The abundances of these species are influenced by their adsorption and desorption on grains.
While no previous literature on BEs was available for ethanolamine and related species, we are attempting for the first time to provide estimated values of BE.
$Ab\ initio$ quantum chemical calculations are carried out using \texttt{Gaussian 09} and \texttt{Gaussian 16} program packages \citep{Frisch2013,fris16} to compute
BEs for ethanolamine and related species having undefined BEs and noted in Table~\ref{table:BE} along with the ground state spin multiplicities.
Among several structural isomers of ethanolamine \citep{nova15}, we consider the lowest-energy structure (g$^\prime$Gg$^\prime$) for all our calculations.
Water ice is the dominant component of the interstellar grain mantle in the dense regions of ISM \citep{boog15}.
Though recent studies \citep{germ22,tina22} suggest that a significant number ($\sim 100$) of H$_2$O molecules in the amorphous solid water (ASW) cluster could simulate the realistic interstellar ice mantle, we assume the water c-tetramer structure to be a unit of  ASW substrate \citep[][showed it gives average percentage deviation from experiment $\sim \pm18.8\%$]{das18} and single binding site despite having larger BE distribution \citep{heyl22} due to a very high computational cost of computing BE distributions.
The BE is typically considered a local property that results from the electronic interaction between the substrate (grain surface or adsorbent) and the species deposited on it (adsorbate). In the case of a bound adsorbate, the BE is a positive value. It is defined as the difference in the electronically optimized energy between the separated species and substrate and the adsorbed species on the substrate.
The optimized energies of the species, water tetramer cluster, and complexes are calculated at the MP2/aug-cc-pVDZ level of theory \citep{dunn89}.
All the BE calculations are performed by optimizing every degree of freedom and using harmonic vibrational zero-point energy (ZPE) corrections to consider molecular vibration at low temperatures. It must be acknowledged that BE calculations are often far from simple approaches like this study and highly depend on the methodology used. \\

\begin{table}
\scriptsize
\centering
\caption{ZPE-corrected BEs considering water tetramer as a substrate using MP2/aug-cc-pVDZ level of theory.
\label{table:BE}}
\begin{tabular}{cccc}
\hline\hline
{\bf Serial} & {\bf Species} & {\bf Ground} & {\bf ZPE-corrected} \\
{\bf Number} &  & {\bf State}  & {\bf Binding Energy (K)} \\
\hline
1 & HNCCO & singlet & 3296 \\
2 & NH$_2$CCO & doublet & 4935 \\
3 & NH$_2$CH & singlet & 6119 \\
4 & CH$_2$OH & doublet & 3864 \\ 
 5 & C$_2$H$_4$OH & doublet &  3656 \\ 
6 & NH$_2$CH$_2$ & doublet & 2987 \\
7 & NH$_2$CHCO & singlet & 2357  \\
8 & NH$_2$CH$_2$CO & doublet & 3044  \\
9 & NH$_2$CH$_2$CHO & singlet & 4314  \\
10 & NH$_2$CH$_2$CH$_2$O & doublet & 3195  \\
11 & $\rm{NH_2CH_2CHOH}$ & doublet & 4509  \\
12 & $\rm{NH_2CH_2CH_2OH}$ & singlet & 4744 \\
\hline\hline
\end{tabular}
\end{table}

\subsubsection{Chemical reaction rates \label{sec:rates}}
\begin{figure*}
\centering
\includegraphics[width=\textwidth]{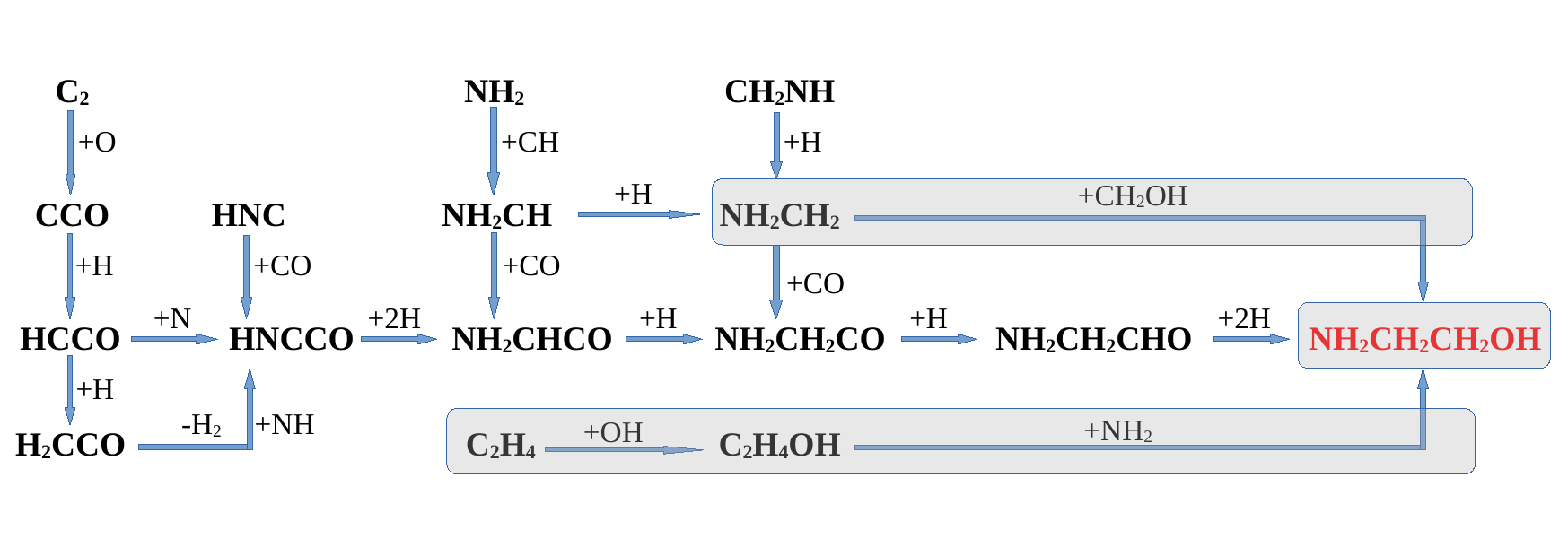}
\caption{Summary of proposed chemical routes leading to the formation of ethanolamine on dust grain surfaces. The grey-shaded area corresponds to the most viable ice-phase routes for the formation of ethanolamine according to the chemical model presented here. Only the formation of ice phase ethanolamine is emphasized here. We note that the destruction by major ions and UV photons (including cosmic-ray-induced ones) is not shown in this diagram, but they are considered in the chemical models.}
\label{fig:reaction_diagram}
\end{figure*}

Table~\ref{table:reaction} and Figure~\ref{fig:reaction_diagram} depicts some ice-phase reactions that have been proposed to lead to the formation of ethanolamine on the surface of dust grains \citep{Rivilla2021,Charnley2001,Wakelam2015,Kameneva2017,Suzuki2018,Krasnokutski2021,ruau16,molp22,sing13,woon02}.
The radical-radical (RR) reactions can happen without any activation barrier, whereas the neutral-radical (NR) reactions have to overcome a finite amount of energy barrier to start. We calculate this ice-phase activation barrier with ZPE correction using the QST2 method by finding a transition state structure with the DFT-B3LYP/6-31+G(d,p) level of theory. Table~\ref{table:reaction} also shows the ice-phase reaction enthalpies (using same level of theory) to check whether the reactions are exothermic or endothermic.
A continuum solvation field is considered to represent the ice features. The integral equation formalism (IEF) variant of the polarizable continuum model (PCM) as a default self-consistent reaction field (SCRF) method is employed with water as a continuous homogeneous dielectric medium representing the solvent \citep{canc97,toma05}. \\

\renewcommand{\baselinestretch}{1.5} 
\begin{table*}
\scriptsize
{\centering
\caption{The ice phase reaction pathways related to the formation of ethanolamine studied using the DFT-B3LYP/6-31+G(d,p) level of theory.}
% The formation of ethanolamine is considered through ice-phase reactions. It has been observed that all these reactions are exothermic. In these reactions, radical-radical reactions can occur without any activation barrier, whereas neutral-radical reactions have to overcome a certain amount of energy barrier to start. The ice-phase activation barriers have been calculated and provided, with zero-point energy (ZPE) correction using the QST2 method to find a transition state structure at the DFT-B3LYP/6-31+G(d,p) level of theory.} 
\label{table:reaction}}
\begin{tabular}{cccccc}
\hline
% {\bf Reaction} & {\bf Reactions} & \multicolumn{3}{c}{\bf Rate coefficient} & {\bf References} \\
% {\bf Number (Type)} &  & $\alpha$ & $\beta$ & $\gamma$ & {\bf and comments} \\
% \hline
% & {\bf Gas-phase reactions} & & & & \\
% \hline
% R1 & $\rm{CCH + OH \rightarrow HCCO + H}$ & $2.2\times10^{-10}$ & 0.0 & 0.0 & KIDA \\
% R2 & $\rm{H + HCCO \rightarrow CH_2 + CO}$ & $1.7\times10^{-10}$ & 0.17 & 0.0 & KIDA \\
\hline
{\bf Reaction} & {\bf Ice-phase/grain-surface} & {\bf Reaction enthalpy}  & {\bf Types of} & {\bf Activation} & {\bf References} \\
{\bf Number (Type)} & {\bf reactions} & {\bf (kcal/mol)} & {\bf reaction} & {\bf barrier (K)} & {\bf and comments} \\
\hline
\hline
R1 (RR) & $\rm{C_2 + O \rightarrow CCO}$ & $-147.46^a$ & Exothermic & ... & KIDA \\
R2 (RR) & $\rm{H + CCO \rightarrow HCCO}$ & $-123.88^a$ & Exothermic & ... & KIDA, \cite{Wakelam2015} \\
R3 (RR) & $\rm{H + HCCO \rightarrow H_2CCO}$ & $-104.18^a$ & Exothermic & ... & KIDA \\
R4 & $\rm{H_2CCO + NH \rightarrow HNCCO + H_2}$ & $-72.72^a$ & Exothermic & & \cite{Rivilla2021} \\
R5 (RR) & $\rm{HCCO + N \rightarrow HNCCO}$ & $-169.75^a$ & Exothermic & ... & \cite{Charnley2001} \\
R6 (NN) & $\rm{HNC + CO\rightarrow HNCCO}$ & $22.60^a$ &  Endothermic$^*$ &  $13289^{a*}$ & \cite{Kameneva2017} \\
\hline
R7a (NR) & $\rm{HNCCO + H \rightarrow NH_2CCO}$ & $-71.99^a$ & Exothermic  & $3000^b$ & \cite{Rivilla2021} \\
R7b (RR) & $\rm{NH_2CCO + H \rightarrow NH_2CHCO}$ & $-77.64^a$ & Exothermic & ... & \cite{Rivilla2021} \\
% R8a (NR) & $\rm{HNCCO + H \rightarrow HNCHCO}$ & $-71.80$ & Exothermic & & \cite{Rivilla2021} \\
% R8b (RR) & $\rm{HNCHCO + H \rightarrow NH_2CHCO}$ & $-77.82$ & Exothermic & ... & \cite{Rivilla2021} \\
\hline
R8 (RR) & $\rm{NH_2 + CH \rightarrow NH_2CH}$ & $-127.64^a$ & Exothermic & ... & \cite{sing13} \\
R9 (RR) & $\rm{NH_2CH + CO\rightarrow NH_2CHCO}$ & $-26.33^a$ & Exothermic & ... & \cite{sing13} \\
% R10 & $\rm{NH_3 + C + CO\rightarrow NH_2CHCO}$ & $-131.69$ & Exothermic & ... & \cite{Krasnokutski2021} \\
\hline
R10a (NR) & $\rm{NH_2CHCO + H \rightarrow NH_2CH_2CO}$ & $-51.17^a$ & Exothermic & $3000^b$ & \cite{Rivilla2021} \\
R10b (RR) & $\rm{NH_2CH_2CO + H\rightarrow NH_2CH_2CHO}$ & $-89.46^a$ & Exothermic & ... & \cite{Rivilla2021} \\
\hline
R11 (RR) & $\rm{NH_2CH + H\rightarrow NH_2CH_2}$ & $-72.15^a$ & Exothermic & ... & \cite{Rivilla2021} \\
R12 (NR) & $\rm{CH_2NH + H\rightarrow NH_2CH_2}$ & $-40.78^a$ & Exothermic & $2134^d$ & \cite{woon02} \\
R13 (NN) & $\rm{NH_2CH_2 + CO\rightarrow NH_2CH_2CO}$ & $-5.35^a$ & Exothermic & $4227^c$ & \cite{sing13} \\
\hline
R14a (NR) & $\rm{NH_2CH_2CHO + H \rightarrow NH_2CH_2CH_2O}$ & $-21.25^a$ & Exothermic & $2369^a$ & \cite{Rivilla2021} \\
R14b (RR) & $\rm{NH_2CH_2CH_2O + H \rightarrow NH_2CH_2CH_2OH}$ & $-102.00^a$ & Exothermic & ... & \cite{Rivilla2021} \\
R15a (NR) & $\rm{NH_2CH_2CHO + H \rightarrow NH_2CH_2CHOH}$ & $-30.85^a$ & Exothermic & $3236^a$ & \cite{Rivilla2021} \\
R15b (RR) & $\rm{NH_2CH_2CHOH + H \rightarrow NH_2CH_2CH_2OH}$ & $-92.40^a$ & Exothermic & ... & \cite{Rivilla2021} \\
\hline
R16 (RR) & $\rm{NH_2CH_2 + CH_2OH\rightarrow NH_2CH_2CH_2OH}$ & $-72.00^a$ & Exothermic & ... & \cite{Rivilla2021} \\
\hline
R17 (NR) & $\rm{C_2H_4 + OH\rightarrow C_2H_4OH}$ & $-27.5^e$ & {Exothermic} & 710$^e$ & \cite{molp22} \\
{R18 (RR)} & $\rm{C_2H_4OH + NH_2\rightarrow NH_2CH_2CH_2OH}$ & {$-79.81$} & {Exothermic} & ... & \cite{molp22} \\
\hline
\hline
\end{tabular} \\

\vskip 0.2cm
$^a$ Calculated in this work. \\
$^*$ Due to the high endothermicity of this reaction, it is not considered in our network. \\
$^b$ Assumed in this work. \\
$^c$ \cite{sing13}. \\
$^d$ \cite{woon02}. \\
$^e$ \cite{molp22}.
\end{table*}
\renewcommand{\baselinestretch}{1}

%These are very helpful for laboratory microwave studies.
%\textbf{ Using these spectroscopic constants, we generate the catalog files in the JPL format using the SPCAT program \citep{Pickett1991}.}

\begin{table*}
\scriptsize
\centering
\caption{Dipole moments using DFT-B3LYP/6-311++G(d,p) level of theory.}
\label{tab:dipole}  
\vskip 0.2cm
\begin{tabular}{ccccc}
\hline
{\bf Species} & \multicolumn{4}{c}{\bf Dipole moments (in D)} \\
\cline{2-5}
 & {\bf $\mu_a$}  & {\bf $\mu_b$} & {\bf $\mu_c$} & {\bf $\mu_{tot}$} \\
\hline
HNCCO  & 1.15 & 0.82 & 0.00 & 1.42 \\
NH$_2$CHCO  & $-2.24$ & $1.41$ & 0.00 & 2.65 \\
$\rm{NH_2CH_2CO}$ & $-2.57$ & $1.45$ & $1.24$ & 3.20 \\
$\rm{NH_2CH_2CH_2OH}$ & $-3.00$ $(2.65)^a$ & 1.14 $(0.89)^a$ & 0.55 $(0.42)^a$  & 3.25 $(2.82)^a$ \\
\hline
\end{tabular} \\
\vskip 0.2cm
{\bf Note:} $^a$ Absolute values of the total dipole moment and its components predicted by \cite{Penn1971}.
\end{table*}

\begin{table*}
\scriptsize
\centering
\caption{Ground vibrational and equilibrium rotational constants and asymmetrically reduced quartic centrifugal distortion constants of HNCCO, NH$_2$CHCO, NH$_2$CH$_2$CO, and $\rm{NH_2CH_2CH_2OH}$ with the DFT-B3LYP/6-311++G(d,p) level of theory. \label{tab:rot_const}}
\begin{tabular}{cccccc}
\hline
{\bf Sl. No.} & {\bf Species}& {\bf Rotational}& {\bf Calculated values}& {\bf Distortion}& {\bf Calculated values} \\
& & {\bf constants} & {\bf (in MHz)}  & {\bf constants} & {\bf (in MHz)} \\
\hline
&& {  ${A_0}$} & 268915.214 & $D_N$& $-0.00039892657$ \\
&& {  ${B_0}$} & 4499.747 &$D_{K}$& $-1034.5240$ \\
1. & HNCCO (Singlet) & {  ${C_0}$} & 4367.401 &$D_{NK}$ & $-2.4075581$    \\
& & {  ${A_e}$} & 219743.385 & $d_N$ & 0.00011340140 \\
&&{  ${B_e}$} & 4476.900 & $d_K$ & $-0.58283144$ \\
&&{  ${C_e}$} & 4387.512 && \\
\hline
&& {  ${A_0}$} & 43671.751 & {  $D_N$} & $-0.0022392657$ \\
&& {  ${B_0}$} & 4608.267 & {  $D_{K}$}& $-0.71283617$ \\
{  2.} & {  $\rm{NH_2CHCO}$ (Singlet)} & {  ${C_0}$} & 4283.586 &{ $D_{NK}$}&    $-0.16315747$\\
& & {  ${A_e}$} & 44190.690 & {  $d_N$} & $-0.00078731244$ \\
&&{  ${B_e}$} & 4634.836 & {  $d_K$} & $-0.015381555$ \\
&&{  ${C_e}$} & 4292.983 && \\
\hline
&& {  ${A_0}$} & 42231.836 & {  $D_N$} & $-0.001274949$ \\
&& {  ${B_0}$} & 4232.048 & {  $D_{K}$} & $-0.73808078$ \\
{  3.} & {  $\rm{NH_2CH_2CO}$ (Doublet)} & {  ${C_0}$} & 3984.143 & {  $D_{NK}$} &  $-0.0060688055$  \\
& & {  ${A_e}$} & 42425.597 & {  $d_N$} & $-0.000071558688$ \\
&&{  ${B_e}$} & 4264.891 & {  $d_K$} & $-0.0090472182$ \\
&&{  ${C_e}$} & 4019.650 && \\
\hline
&& {  ${A_0}$}& 14401.029  & $D_N$& $-0.0061629178$  \\
&& {  ${B_0}$}& 5429.504 &$D_{K}$ & $-0.048060075$ \\
4. & $\rm{NH_2CH_2CH_2OH}$ (Singlet) & {  ${C_0}$} & 4479.870 &$D_{NK}$ & $0.020970813$ \\
& & {  ${A_e}$} & 14460.409 ($14508.73\pm0.1^a$) & $d_N$ & $-0.0018253369$  \\
&& {  ${B_e}$} & 5538.498 ($5546.46\pm0.04^a$)& $d_K$ & $-0.011234133$  \\
&&{  ${C_e}$} & 4552.708 ($4570.46\pm0.03^a$) & & \\
\hline
\hline
\end{tabular} \\
\vskip 0.2cm
{\bf Note:} $^a$ \cite{Penn1971}
\end{table*}

%\subsubsection{Vibrational spectroscopy}
%\textbf{ NOT NEEDED THIS SECTION!} We perform IR calculation of ethanolamine using the DFT-B3LYP/6-311G(d,p) level of theory. We employ the integral equation formalism (IEF) variant of the polarizable continuum model (PCM) as a default self-consistent reaction field (SCRF) method with water as a solvent.

\begin{table*}
\caption{Comparing IR peak positions in the ice phase (from this work) with that of the previously available in \cite[pure liquid phase experiment and theoretical gas phase]{Jackson2009} and \cite[in C$_2$Cl$_4$ solution]{Krueger1965}.}
	\begin{tabular}{cc|ccc}
		\hline
		\multicolumn{5}{c}{Peak position(cm$^{-1}$)}                                                  \\ \hline
		\multicolumn{2}{c|}{This work}               & \multicolumn{3}{c}{Previous works}                   \\ \hline
		\multicolumn{2}{c|}{Experiment}  & Pure Liquid phase & in C\textsubscript{2}Cl\textsubscript{4} solution & Theoretical gas phase \\ \cline{1-2}
		7 K           & 180 K                                        &    \cite{Jackson2009}                                &          \cite{Krueger1965}                          &     \cite{Jackson2009}                                   \\ \hline
		            &                              &                               & 3640                               & 3644                                   \\
		           &                              &                               & 3624                               & 3612                                   \\
		           &                              &                                    & 3526                               & 3508                                   \\
		            &                              &                                    & 3411                               &                                   \\
		3352          & 3314                                         & 3355                               & 3340                              &                                   \\
		3287          & 3268                                         & 3293                               &                              &                                    \\
		3201          & 3182                                        &                                    &                               &                                   \\
		& 3136                                         &                                    &                                & 3036                                   \\
		& 3073                                         &                                    &                                & 2995                                   \\
		2936          & 2911                                         & 2927                               &                                    &  2956                                  \\
		2861          & 2874                                         & 2864                               &                                    &  2948                                  \\
		2786          & 2768                                         &                                    &                                    &                                    \\
		2726          & 2735                                         &                                    &                                    &                                    \\
		2658          & 2670                                         &                                    &                                    &                                   \\
		& 2642                                         &                                    &                                    &                                   \\
		& 2567                                         &                                    &                                    &                                    \\
		& 2513                                         &                                    &                                    &                                     \\
		1684          & 1674                                         &                                    &                                    &  1672                                      \\
		1606          & 1615                                         & 1601                               &                                    &                                        \\
		1460          & 1489                                         & 1461                               &                                    &  1465                                      \\
		& 1452                                         &                                    &                                    &                                        \\
		1395          & 1385                                         &                                    &                                    &  1405                                      \\
		1360          & 1362                                         & 1355                               &                                    &                                        \\
		& 1348                                         &                                    &                                    &                                        \\
		& 1310                                         &                                    &                                    &                                        \\
		& 1294                                         &                                    &                                    &                                        \\
		1254          & 1257                                         &                                    &                                    &                                        \\
		1174          & 1160                                         & 1172                               &                                    &  1198                                       \\
		1083          & 1119                                         & 1081                               &                                    &                                        \\
		& 1101                                         &                                    &                                    &  1101                                       \\
		& 1066                                         &                                    &                                    &  1060                                      \\
		1035          & 1030                                         & 1033                               &                                    &                                        \\
		& 986                                          &                                    &                                    &        999                                  \\
		970           & 962                                          &                                    &                                    &                                      \\
		872           & 874                                          &                                    &                                    &                                        \\
		& 839                                          &                                    &                                    &                                        \\
		& 774                                          &                                    &                                    &                                        \\
		\hline
	\end{tabular}
		\label{tab:ir}
\end{table*}

\subsubsection{Chemical modeling}
The CMMC (Chemical Model of Molecular Cloud) code \citep{das15,das19,das21,gora17a,gora17b,gora20,sil18,sil21,bhat21,sriv22,Tani23} is implemented for studying the formation of
ethanolamine.
It is a time-dependent gas-grain code that considers gas and homogeneous reactive ice mantle, i.e., a two-phase model, based on the rate equation approach.

Here, the gas-phase pathways %\st{of the CMMC code} 
are mainly adopted from the UMIST Database for Astrochemistry 2022 \citep{umis22}.
%\textbf{ [Can we use UMIST 2023 updated network (\url{https://umistdatabase.uk/downloads})?]}.
Additionally, the destruction of gas phase ethanolamine and its related species by major ions (${\rm H_3}^+$, HCO$^+$, H$^+$, C$^+$, He$^+$ with a typical rate constant of $\sim 10^{-9}$ cm$^3$ s$^{-1}$) and photo reactions are considered (see Table~\ref{sec:dest}). Altogether, our gas phase network contains $\sim 8848$ reactions between $\sim 750$ species.
A typical cosmic-ray ionization rate of the ISM $\sim 1.3 \times 10^{-17}$~s$^{-1}$ is considered. %\st{in our model}. \st{The BE ($E_d$) of the surface species plays a decisive role in controlling the chemical complexity of interstellar ice. A straightforward relation between the diffusion energy ($E_b$) of a species with the BE is considered by following $E_b=R E_d$, where $R$ may vary between 0.35 and 0.8} \citep{garr07}. %\st{Here, we use a typical value $R=0.5$.} 
A typical ratio between the desorption and diffusion energy of $\sim 0.5$ is considered.
Here, the surface chemistry network is primarily considered from \cite{ruau16}, and the BE of the surface species is considered from \cite{garr13,das18,sriv22,mond21,wake17,sil21}. %\cite{ruau16} and \cite{das18}.
The BE set used in this work is available at \url{https://www.iases.org.in/Ankan-web/AD-doc/BE-astrochem-kol-ver1.dat}.
Additionally, the ice phase reaction network contains other pathways used in \cite{garr17,bell14}
regarding the H abstraction reactions. Moreover, we incorporate the ice phase reactions detailed in Table~\ref{table:reaction} and Figure~\ref{fig:reaction_diagram}
and the %\st{binding energy} 
BE reported in Table~\ref{table:BE}. Altogether our present surface network contains 1485 reactions and $\sim 500$ ice phase species. Various ice phase desorption mechanisms (thermal, cosmic ray induced, reactive, and photodesorption) have been considered for the desorption of the newly included species.
%\st{The ice-phase pathways listed in Table} \ref{table:reaction} 
%\st{are also considered in our network. We have used the reaction rates calculated in Section} \ref{sec:rates} \st{and noted in Table} \ref{table:reaction}.
% For the newly included species, cosmic ray-induced dissociation, photo-dissociation, and ion-neutral destruction by some of the major ions (${\rm H_3}^+$, HCO$^+$, H$^+$, C$^+$, He$^+$ \textbf{ with typical rate constant $\sim 10^{-9}$ cm$^3$ s$^{-1}$}) in the gas phase are considered with typical rate constant $\sim 10^{-9}$ cm$^3$ s$^{-1}$.
The encounter desorption \citep{hinc15,chang20,das21} effect is considered to avoid overestimating the H$_2$ abundance on the grain surface.
%A reactive desorption with the fiducial parameter of $a=0.01$ is considered for all the barrierless reactions forming a single product on grain.
Recently, \cite{oba19} and \cite{furu22} estimated the reactive desorption probability of H$_2$S and PH$_3$  $\sim 3-4$ \%.
However, due to our limited understanding of this parameter, a typical value $\sim 0.01$ is adopted \citep{garr13,furu15,das21}.
%\st{ The low metallic elemental abundances are used as the initial abundances} \citep[EA1 set]{wake08}.
Typically, low metal abundances are used for this type of model. Currently, we are utilizing the values used by \cite{garr13} for a similar model, which is based on the initial low metal elemental abundance set by \cite{grae82}, except for He, C$^+$, N, and O, for which values were derived from their diffuse cloud model. In this context, initial H, H$_2$, He, N, and O are considered neutral, while all other species are in their ionized form. The initial electron abundance is set equal to the total initial elemental abundances of positive ions.

\begin{figure*}
\centering
\includegraphics[width=8.5cm,height=7cm,angle=0]{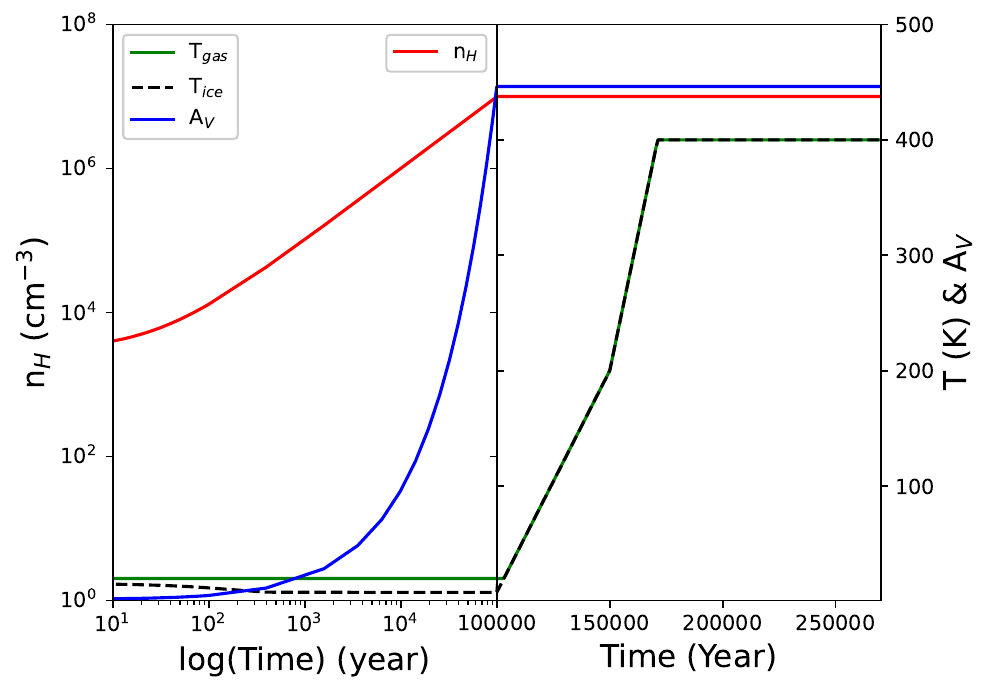}
\includegraphics[width=8.5cm,height=7cm,angle=0]{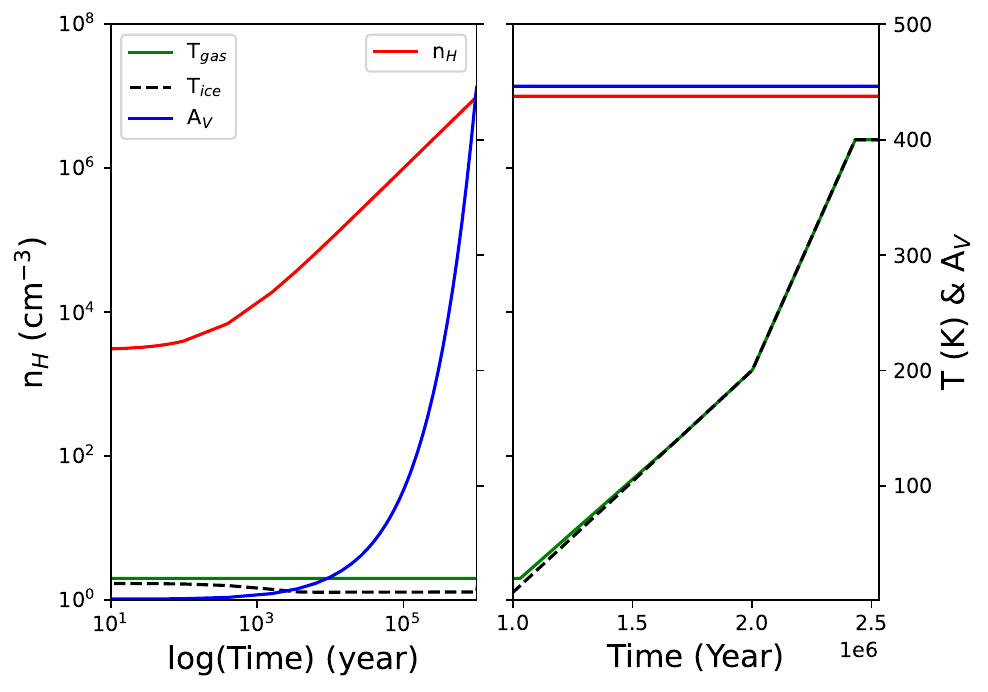}
\caption{The adopted physical conditions for a hot core (left panel) and hot corino (right panel) region.}
\label{fig:phys}
\end{figure*}

\begin{table*}[]
    \centering
    \caption{Adopted physical parameters considered in our model.}
        \begin{tabular}{|c|cc|}
    \hline
         &{hot core}& {hot corino}  \\
         \hline
         &\multicolumn{2}{c|}{Collapse phase}\\
         \cline{2-3}
         ${\rm t_{coll} (yr)}$& $10^5$& $10^6$\\
         ${\rm {n_H}_{min} (cm^{-3})}$&$3.0 \times 10^3$&$3.0 \times 10^3$\\
         ${\rm {n_H}_{max} (cm^{-3})}$&$10^7$&$10^7$\\
         ${\rm Tgas (K)}$& 20  & 20\\
         $A_V$&$\rm{A_{V0}(n_H/{n_H}_{min})^{2/3}}$($A_{V0}=2$)&$\rm{A_{V0}(n_H/{n_H}_{min})^{2/3}}$($A_{V0}=2$)\\
         ${\rm Tice (K)}$&\cite{zucc01,garr08}&\cite{zucc01,garr08}\\
         \hline
         &\multicolumn{2}{c|}{1$^{st}$ warm-up phase}\\
         \cline{2-3}
         ${\rm t_{warm1} (yr)}$&$5.0 \times 10^4$&$10^6$\\
         ${\rm {n_H} (cm^{-3})}$&$10^7$&$10^7$\\
        ${\rm {Tgas_{min}} (K)}$&20&20\\
         ${\rm {Tgas_{max}/Tice_{max}}^b (K)}$&200&200\\
        ${\rm {Tice_{min}}} (K)$&${\rm {Tice_{min}}^a}$&${\rm {Tice_{min}}^a}$\\
        ${\rm A_V}$&${A_V}_{max}^a$&${A_V}_{max}^a$\\
         \hline
         &\multicolumn{2}{c|}{2$^{nd}$ warm-up phase}\\
         \cline{2-3}
         ${\rm t_{warm2} (yr)}$&$2.1 \times 10^4$&$4.3 \times 10^5$\\
         ${\rm {n_H} (cm^{-3})}$&$10^7$&$10^7$\\
         ${\rm {Tgas_{min}/Tice_{min}} (K)}$&200&200\\
         ${\rm {Tgas_{max}/Tice_{max}}^b (K)}$&400&400\\
          ${\rm A_V}$&${{A_V}_{max}}^a$&${{A_V}_{max}}^a$\\
        \hline
         &\multicolumn{2}{c|}{post-warm-up phase}\\
         \cline{2-3}
${\rm t_{pw} (yr)}$&$10^5$&$10^5$\\
         ${\rm {n_H} (cm^{-3})}$&$10^7$&$10^7$\\
         ${\rm {Tgas} (K)}$&400&400\\
        ${\rm {Tice} (K)}$&400&400\\
        ${\rm A_V}$&${{A_V}_{max}}^a$&${{A_V}_{max}}^a$\\
                 \hline
       Total simulation time &$2.712 \times 10^5$&$2.53 \times 10^6$\\
        \hline
    \end{tabular} \\
    \vskip 0.2cm
{\bf Notes:} $^a$ Obtained at the end of the collapsing phase. \\
$^b$ A linear relationship between temperature increase and time is being considered. \\
    \label{tab:time}
\end{table*}

%\subsubsection{Physical condition}
Here, a physical condition is used, which is suitable to explain the environment corresponding to the surroundings of high-mass and low-mass protostars, i.e., hot core and hot corino, respectively. The physical condition of our model is divided into three distinct stages: %\st{isothermal} 
collapse, warm-up, and post-warm-up.
During the warm-up stage, the collapse halts, and dust and gas temperatures gradually increase. The time scale of this phase is primarily based on the approach of \cite{viti04}, who used the observed protostellar luminosity function of \cite{moli00} to derive effective temperatures throughout the accretion phase of a central protostar until it reaches Zero age main sequence. They approximated the compelling temperature profile as a power law concerning the protostar's age. They used contraction times determined by \cite{bern96} to estimate a timescale of $10^4 - 10^6$ years for stars of 60 to 5 solar masses, respectively.
The physical parameters adopted in our model are summarized in Table~\ref{tab:time} and Figure~\ref{fig:phys}.
Similar physical parameters were adopted in some of the earlier studies \citep{garr06,garr13,garr17,gora20,mond21,sriv22,Tani23}.

\subsubsection{Microwave rotational spectroscopy}
Several of the proposed precursors of ethanolamine have not had their rotational spectroscopy measured (see Figure~\ref{fig:reaction_diagram}). Therefore, we are presenting the theoretical calculations of its dipole moments and rotational constants to aid future laboratory experiments.
Rotational spectroscopy is the primary tool to detect species in the ISM. Almost $80\%$ of the observed species having permanent electric dipole moments are detected from their rotational transitions \citep{mcgu22}. The intensity of any rotational transition is directly and inversely proportional to the square of the dipole moment and the rotational partition function, respectively. 
Dipole moment components along the inertial axis and total dipole moments of ethanolamine and related molecules (HNCCO, NH$_2$CHCO, and NH$_2$CH$_2$CO, for which no data were available in the literature) calculated using DFT-B3LYP/6-311++G(d,p) level of theory are summarized in Table~\ref{tab:dipole}.
In accordance with the experimental work by \cite{Penn1971}, a very strong a-type dipole moment for ethanolamine is obtained, while the b-type and c-type lines are generally much weaker, indicating that ethanolamine has a predominant dipole moment along one principal axis, which has implications for its spectroscopic properties and structural characteristics.
However, we note some deviations in the dipole moments compared to the values reported by \cite{Penn1971}. This suggests that employing a high-cost functional and basis set combination could help reduce these deviations, which is beyond the scope of this work.
The values obtained by \cite{Penn1971} for ethanolamine are based on experiments and could be considered more accurate.
Our calculated dipole moments for HNCCO, NH$_2$CHCO, and NH$_2$CH$_2$CO are relatively high (see Table \ref{tab:dipole}), which would help to detect the molecules once the rotational spectroscopy is measured in the future.

The ground vibrational ($A_0$, $B_0$, $C_0$) and equilibrium ($A_e$, $B_e$, $C_e$) values of rotational constants and asymmetrically reduced
quartic centrifugal distortion constants ($D_N$, $D_K$, $D_{NK}$, $d_N$, $d_K$) are also calculated for these species using the same level of theory and noted in Table~\ref{tab:rot_const}.
%\sout{Table~\ref{tab:rot_const} shows that our calculated parameters for ethanolamine align with those obtained by \cite{Penn1971}, indicating the reliability of the data.}
The percentage deviations of the rotational constants A, B, and C calculated in our work compared to those in \cite{Penn1971} are underpredicted by $0.33\%$, $0.14\%$, and $0.39\%$, respectively.
These small percentage deviations indicate a high degree of consistency between the two sets of results.
Based on this, we provide the rotational spectroscopic constants for its related species, HNCCO, NH$_2$CHCO, and NH$_2$CH$_2$CO, for which no values were available in the literature.

\section{Results and discussion} \label{sec:result}

\subsection{Experimental results}

\subsubsection{VUV spectra of pure ethanolamine ice}
\begin{figure}
	\centering
	 \includegraphics[width=0.5\textwidth]{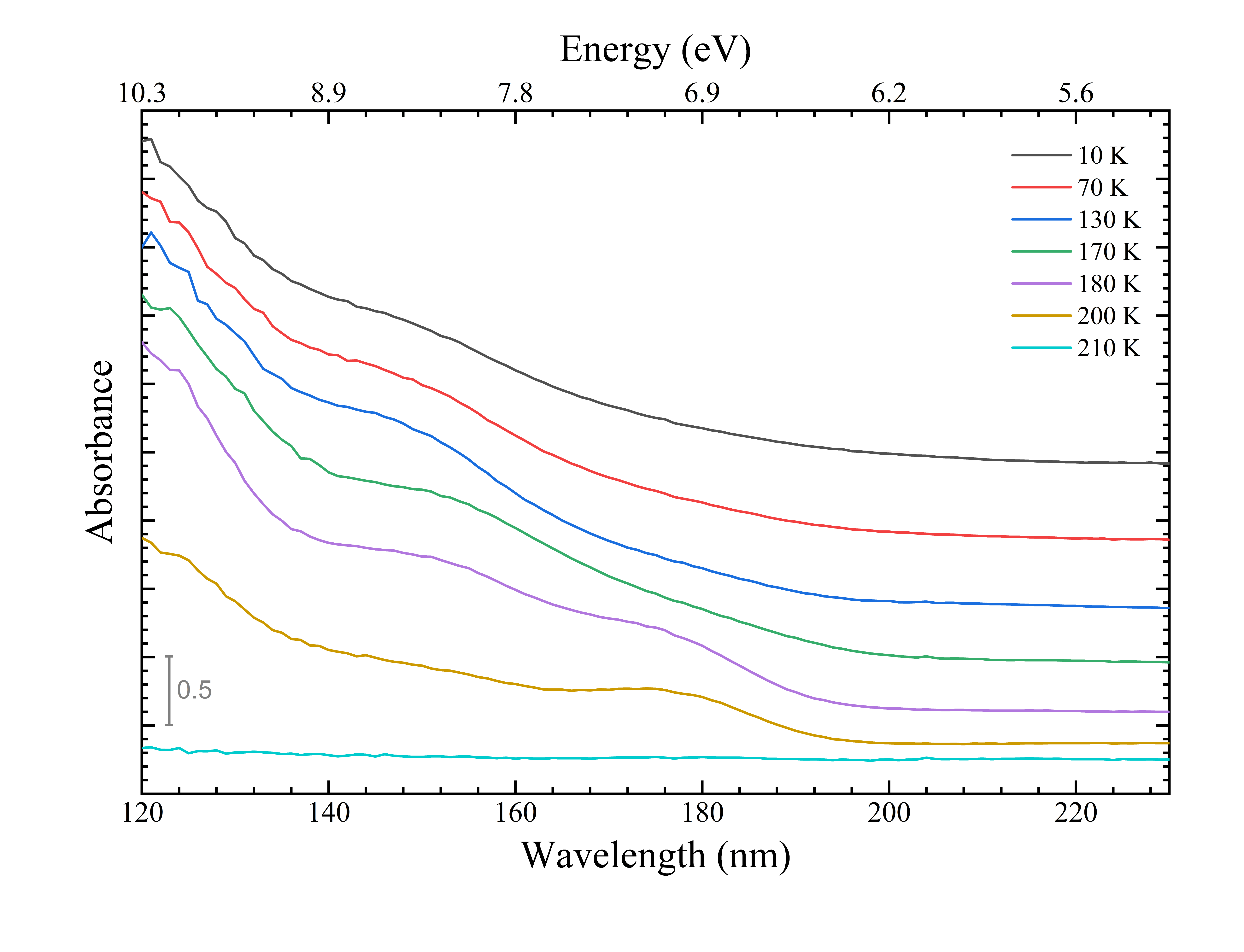}
	\caption{Experimental VUV spectra of ethanolamine ices (stacked) in the 120 - 240 nm range for a selected set of temperatures}.
	\label{vuvstack}
\end{figure}

Figure~\ref{vuvstack} presents the stacked temperature-dependent VUV spectra of ethanolamine ice deposited at 10~K in the $120-230$~nm range and subsequently warmed to higher temperatures. 

At 10~K, we observe that the spectrum has a weak band/bump between 141 and 157~nm and beyond 210~nm, there is no absorption. When warmed to higher temperatures, the spectra show no noticeable changes in the spectra until 100~K. On warming the ice to 130~K, we could observe that the band centered around 142~nm became more prominent while the peak positions remained unchanged. 
 On further warming to 180~K, the spectrum peaked at 172 nm. At 180~K, the infrared red spectrum reveals an onset phase change. The two distinct bands could be due to the dimer, monomer forms of the ethanolamine, as mentioned earlier in the case of formamide ice \citep{sivaraman2012}. Upon warming to 200~K, spectra recorded clearly show a completely crystalline phase, and beyond 210~K, the ice sublimes.

\subsubsection{IR spectra of pure ethanolamine ice}
\begin{figure}
	\centering
	\includegraphics[width=0.5\textwidth]{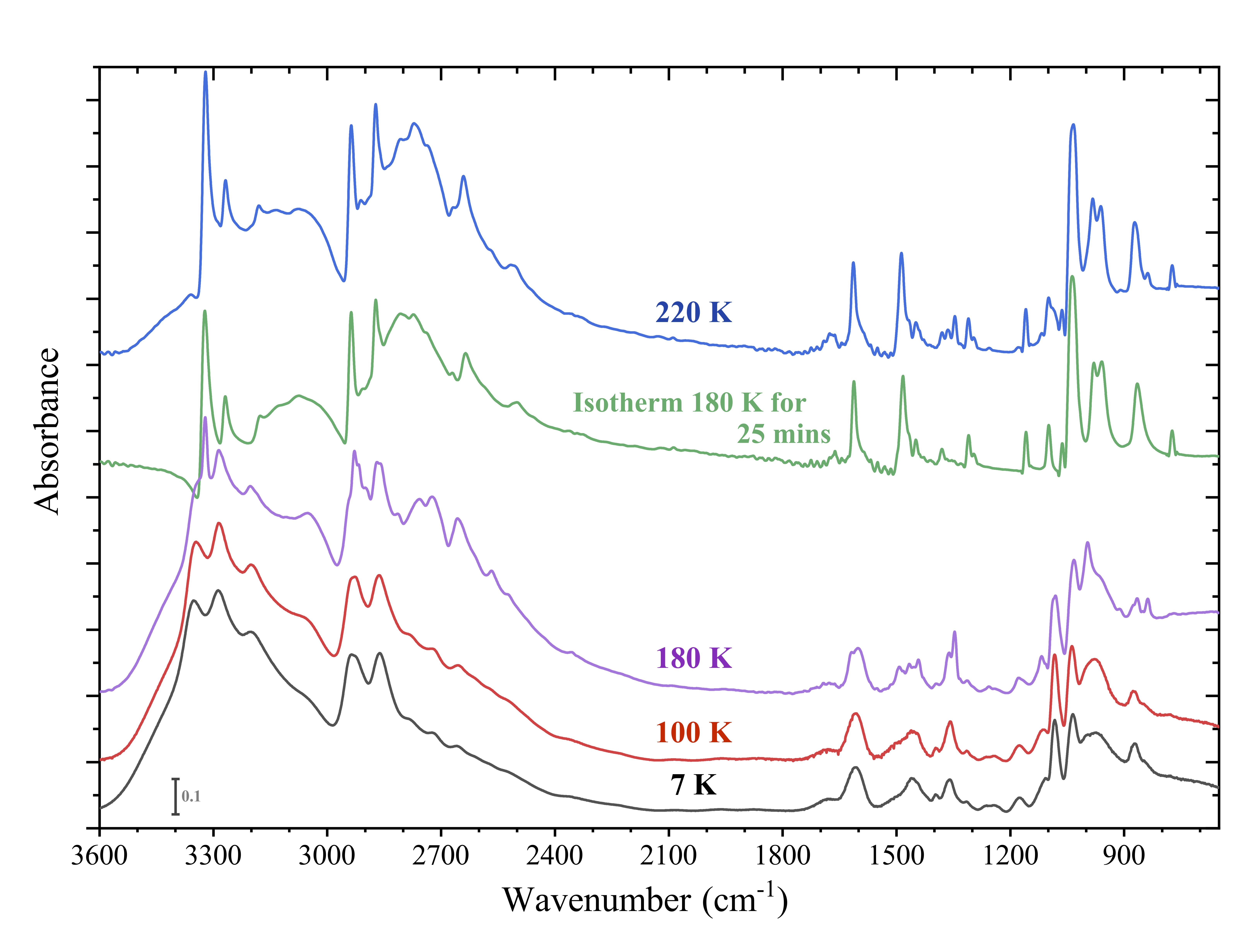}
	\caption{Temperature dependent experimental IR spectra of ethanolamine ices.}
	\label{fig:ir}
\end{figure}

Figure~\ref{fig:ir} depicts the temperature-dependent stacked IR spectra of ethanolamine ice deposited at 7~K between 3600~cm$^{-1}$ and 750~cm$^{-1}$. When deposited at 7~K, several peaks are noticed in two regions, one between 3600~cm$^{-1}$ and 2500~cm$^{-1}$ and the other between 1700~cm$^{-1}$ and 800~cm$^{-1}$. The 1$^{st}$ column of Table~\ref{tab:ir} presents the positions of all these observed peaks. In the absence of any previous ice-phase spectra, comparing this with the previous liquid phase and gas-phase results, the peak positions of pure ethanolamine at 10~K are in good agreement with that of the pure liquid phase of \cite{Jackson2009}, and also with their theoretical gas-phase values. We see no significant difference up to 180~K after heating to higher temperatures. At 180~K, we see clear evidence of compaction of the ethanolamine ice, and when kept at 180~K for about 25~minutes, the phase change (amorphous/metastable to crystalline) is noticed. When this spectrum (isotherm at 180~K for 25~min spectrum of Fig~\ref{fig:ir}) is compared with other spectra at lower temperatures, many additional peaks and sharper peaks are noticed. These are a result of the reorientation of ethanolamine molecules in the ice. The ice wholly sublimated at 230~K. 
 
\par
\begin{figure}
	\centering
	\includegraphics[width=0.5\textwidth]{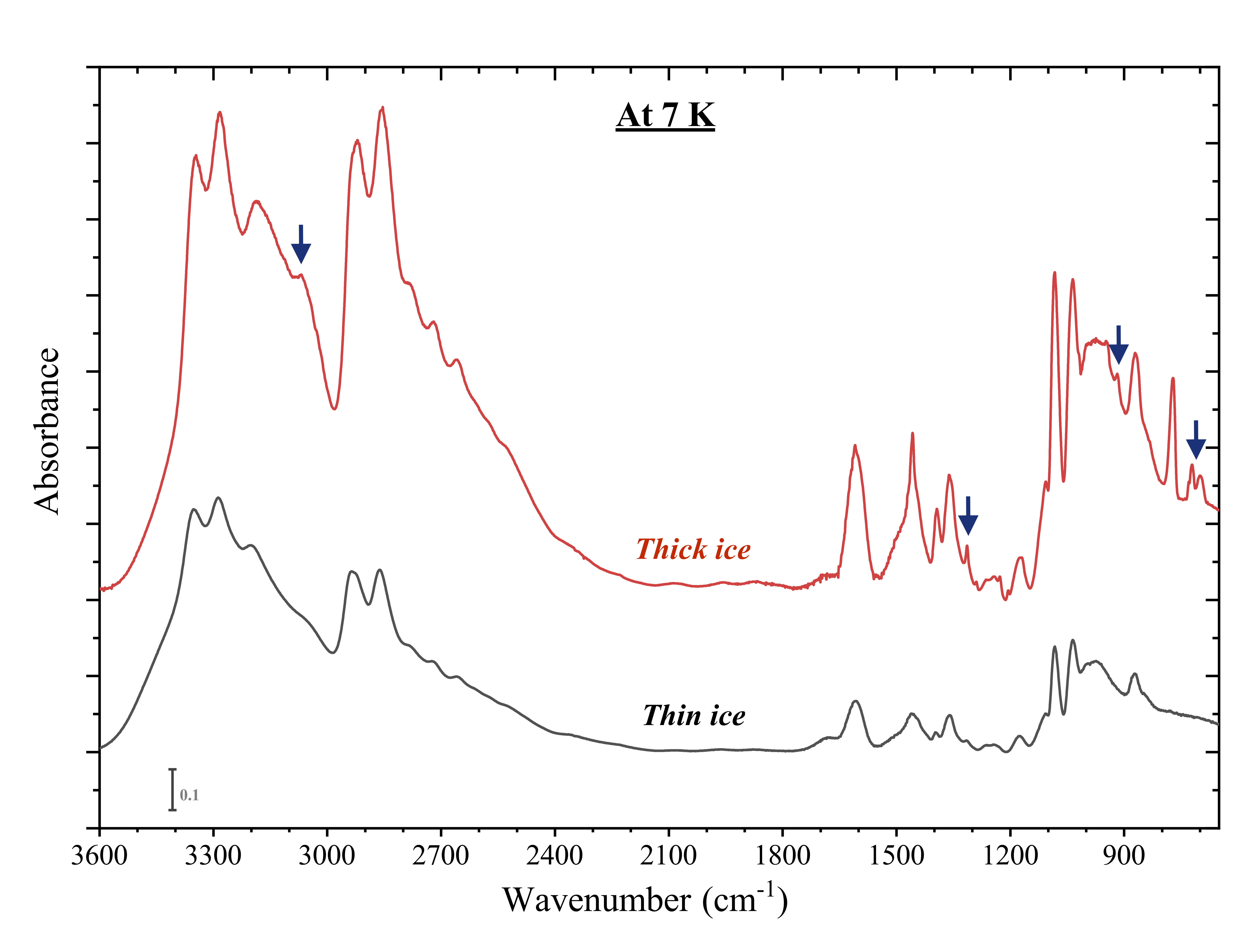}
	  \includegraphics[width=0.5\textwidth]{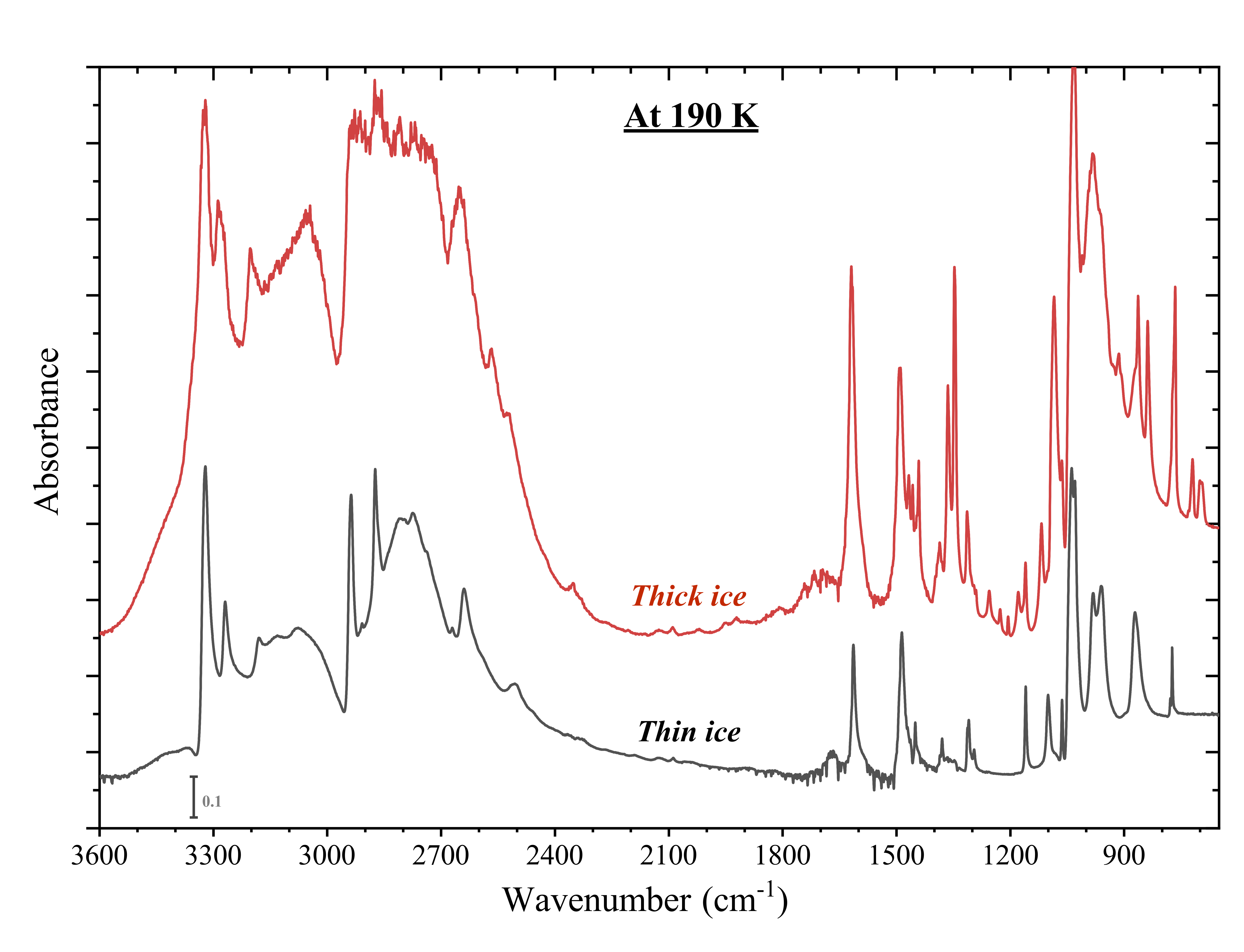}
	\caption{Comparison of Experimental IR spectra of Thin and Thick ethanolamine ices at 7~K (upper panel) and after warming to 190~K (lower panel). The additional peaks in thick ice appearing at 7~K have been marked with blue arrows.}
	\label{fig:comp}
\end{figure}

 Another exciting aspect of ethanolamine ice is its dependence on the thickness of the ice. %\st{As we can notice from the upper panel of Fig.} \ref{fig:comp} \st{that a thick layer of ethanolamine is deposited at 7 K, several additional peaks are noticed compared to that of thin ice.}
The upper panel of Figure~\ref{fig:comp} illustrates the deposition of ethanolamine ice at 7~K, while the lower panel displays it for 190~K. The observation indicates that at the lower deposition temperature, several additional peaks (indicated by arrows in the upper panel of Figure~\ref{fig:comp}) are present for the thick ice, which were absent in the thin ice.

A more detailed and quantitive study is needed to understand the origin and 
%\st{assingments} 
assignments for these peaks, nonetheless, it is a noteworthy observation. At 190 K (lower panel of Figure~\ref{fig:comp}), evidence of phase change is noticed in almost all the new peaks 
%\st{noticed}
in the thick ice at 7 K. Since some regions are already saturated, all the peaks are not seen. To further understand the behavior of ethanolamine ices, we performed several cycles of experiments by changing different variables like deposition temperature, deposition time, isotherm temperature, and isotherm time, a few results of which are shown in Table~\ref{tab:expt}. From a series of experiments summarised in Table~\ref{tab:expt}, we can infer that irrespective of the deposition time or rate of warming, the phase change temperature remains the same (190~K) when the ethanolamine ice is deposited at 7~K. The same ice deposited at 7~K, when kept at isotherm at 180~K for $\sim$ 4~hrs, it changes from amorphous to crystalline at 180~K itself. But when kept at isotherm at 170~K, the phase change does not occur even after 7~hrs. This shows that the phase change process initiates around 175~K, but it is a slow process and hence requires more energy, which it could either get from higher thermal energy (higher temperature) or cumulative energy due to isotherm at any temperature above 175~K.
  
\begin{table*}
 \centering
 \caption{Log of IR experiments carried out at different temperatures and thicknesses and different isotherm temperatures and durations.
 \label{tab:expt}}
 \begin{tabular}{ccccccc}
 \hline
 {\bf Dep.} & {\bf Dep.} & {\bf Phase} & {\bf Ramp} & {\bf Sublimation} & {\bf Isothermal} & {\bf Remarks} \\
 {\bf temp.} & {\bf time} & {\bf change} & {\bf rate} &  & {\bf temp. and} &  \\
 {\bf (K)} & {\bf (sec)} & {\bf (K)} & {\bf (K/min)} & {\bf (K)} & {\bf duration} & \\
 \hline
 7 & 220 & 190 & 10 & $235-240$ & --- & --- \\
 7 & 903 & 190 & 5 & 230 & --- & --- \\
 7 & 425 & 190 & 5 & 230 & --- & --- \\
 7 & 435 & 180 & 5 & 230 & 180~K, 4~hrs & Crystalline within 4 hrs of isothermal at 180 K \\
 170 & 2400 & 190 & 5 & 230 & 170~K, 7~hrs & Compaction after 2 hrs, No crystallization even after 7 hrs. \\
 175 & 1500 & 175 & 5 & 220 & 175~K, 90~min & The ice turned crystalline \\
 180 & 1200 & 180 & 5 & 230 & --- & Fully crystalline ice \\
 \hline
 \end{tabular}
\end{table*}

\begin{figure}
\centering
\includegraphics[width=0.5\textwidth]{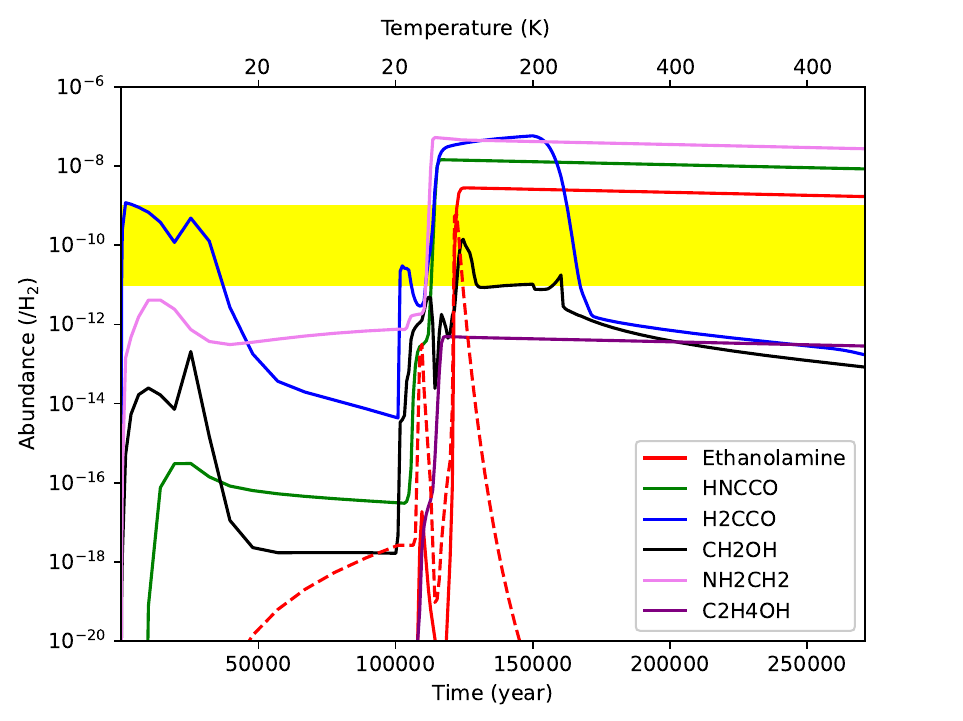}
%\vskip 0.5cm
\includegraphics[width=0.5\textwidth]{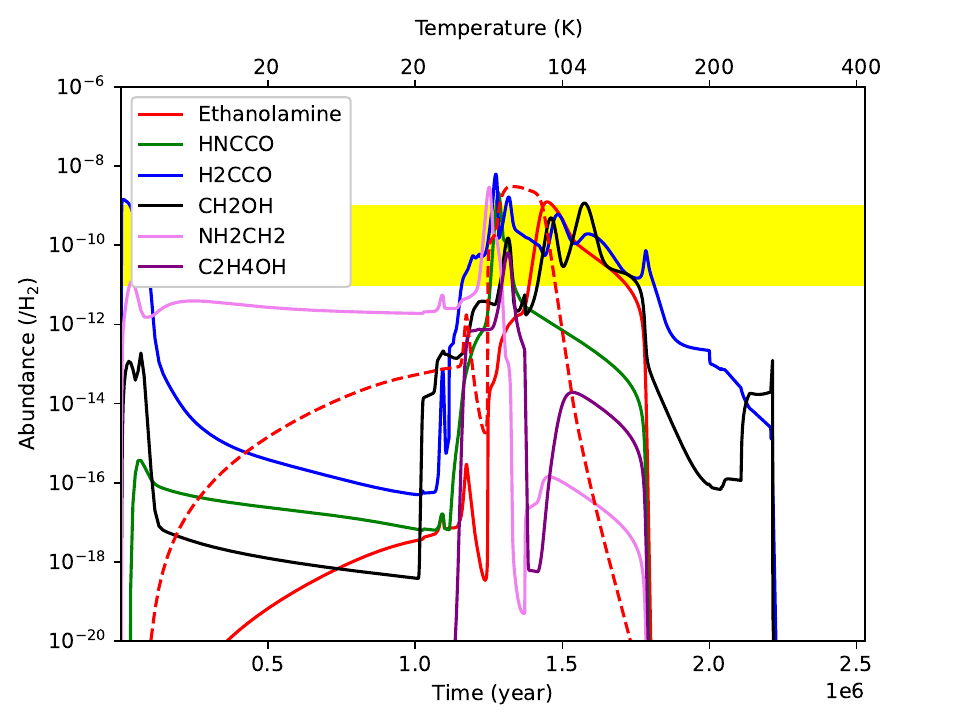}
\caption{Time evolution of the ethanolamine abundances and some related species (HNCCO, H$_2$CCO, CH$_2$OH, NH$_2$CH$_2$, $\rm{C_2H_4OH}$) for hot core (upper panel) and hot corino region (lower panel). The dashed line represents the ice phase ethanolamine abundance %\st{in the absence of the reaction R16}
.}
\label{fig:abund-case}
\end{figure}

\subsection{Computational results}

\subsubsection{Chemical modeling of hot core}
%\st{these}
Our quantum chemical calculations have revealed that all the reactions mentioned in Table~\ref{table:reaction} are exothermic. Among these reactions, the neutral-radical reactions possess activation barriers. These activation barriers have been estimated and directly incorporated into our model. In addition, some other reactions are presented based on our study. 
For the ice-phase synthesis of H$_2$CCO, the barrierless reactions R1-R3 are considered from KIDA.
\cite{Charnley2001} proposed the formation of HNCCO by the reaction between HCCO and N (reaction R5).
This radical-radical (RR) reaction can happen at each encounter without any activation barrier. It is noticed that this reaction is exothermic, having reaction enthalpy $-169.75$~kcal/mol.
{\cite{Rivilla2021} proposed reaction R4 for the formation of HNCCO.}
\cite{Kameneva2017} suggested the neutral-neutral (NN) reaction between HNC and CO (reaction R6). However, our calculation exhibits it as an endothermic reaction ($22.60$~kcal/mol). Also, with the transition state calculation, it is noticed that it possesses a very high activation barrier ($\sim 13289$~K).
In our reaction network, reaction R6 is not considered for forming HNCCO. 
Two successive hydrogenations of HNCCO are required to produce $\rm{NH_2CHCO}$. The first step (R7a) is the neutral-radical (NR) reaction, and the second step (R7b) is the RR reaction. Both reactions are found to be exothermic.
Since the second reaction is RR, it will process as barrierless. But the first reaction should contain some activation barrier. However, an actual transition state of the reaction R7a is not obtained. Therefore, a conservative value of $3000$~K is considered an educated estimation for the activation barrier of this reaction. The effect of this barrier is also discussed afterward.

There are other alternative pathways to form $\rm{NH_2CHCO}$ by the reactions R8 and R9 \citep{sing13}. Both these RR reactions are barrierless and exothermic. Therefore, these pathways are also considered in our network.  NH$_2$CHCO is further processed through two successive hydrogenations (R10a and R10b) to form NH$_2$CH$_2$CHO.
The first reaction (R10a) should contain the activation barrier. However, we could not get an actual transition state and consider the $3000$~K barrier for R10a. At the same time, the second RR reaction, R10b, is barrierless and can be processed without an activation barrier.

The reactions R11-R13 provide another alternative formation pathway for $\rm{NH_2CH_2CO}$, where R11 is RR barrierless exothermic reaction and R12 is NR exothermic possessing a barrier of 2134~K \citep{woon02}. Following \cite{sing13}, we consider the ice-phase NN reaction R13 with an activation barrier 4227~K.
Hydrogenation of $\rm{NH_2CH_2CHO}$ can happen in two possible locations (C or O atom positions) and form $\rm{NH_2CH_2CH_2O}$ (R14a) or NH$_2$CH$_2$CHOH (R15a). Quantum chemical calculations yield activation barriers of 2369~K and 3236~K for R14a and R15a, respectively.
$\rm{NH_2CH_2CH_2O}$ and $\rm{NH_2CH_2CHOH}$ further react with another hydrogen atom and form the target product ethanolamine (reactions R14b and R15b). Both of these RR reactions are found to be exothermic. The exothermic reaction R16 between two radicals $\rm{NH_2CH_2}$ and $\rm{CH_2OH}$ to form ethanolamine as proposed by \cite{Rivilla2021} are included in our network.
Recently, \cite{molp22} proposed two subsequent surface reactions in forming ethanolamine by $\rm{C_2H_4+OH \rightarrow C_2H_4OH}$ (R17, NR reaction) and $\rm{C_2H_4OH+NH_2 \rightarrow NH_2CH_2CH_2OH}$ (R18, RR reaction).
Furthermore, we consider an additional ice phase barrier less reaction between
$\rm{NH_2CH_2CH_2OH}$ and HCOOH
for the formation of $\beta-$alanine.

Figure~\ref{fig:abund-case} shows the abundances of ethanolamine along with some of its related species (HNCCO, H$_2$CCO, CH$_2$OH, NH$_2$CH$_2$, and C$_2$H$_4$OH) obtained from our model. The lower X-axis in Figure~\ref{fig:abund-case} shows the time, the upper X-axis shows gas temperature, and the Y-axis represents the abundance with respect to H$_2$.
The upper and lower panels represent the time evolution of abundances for hot core and hot corino regions, respectively.
Dashed lines show the abundance of ethanolamine in the absence of the reaction R16. Since the hydrogenation reaction by physisorption is not generally efficient beyond 20~K, we find a significant contribution of ethanolamine by the surface reaction between CH$_2$OH and NH$_2$CH$_2$ (reaction R16).
In the absence of the reaction number R16, ethanolamine formation is dominated by another surface reaction between $\rm{C_2H_4OH}$ and NH$_2$ (reaction number R18).

In the case of a hot core, we consider a fast collapsing and a fast warm-up time scale. The upper panel of Figure~\ref{fig:abund-case} shows that a peak abundance of $\sim 2.8 \times 10^{-9}$  
%\st{(considering reaction R16 and R18)} 
could be achieved around 
%\st{$90$~K} 
$100$~K for a hot core region. 
%\st{The absence of reaction number R16 (dashed line) shows a peak abundance of $1.9 \times 10^{-11}$.}
We notice that most of the ice phase formation of ethanolamine happens here between 
%\st{$40-60$~K.} 
$60-90$~K (red dashed line in Figure~\ref{fig:abund-case} depicts ice phase ethanolamine). Due to a shorter warm-up time scale, its destruction would not have taken place for a long period; as a result, we can see a healthy abundance of ethanolamine at the end of the simulation. On the contrary, we consider a slow collapsing and warm-up time scale for the hot corino (lower panel of Figure~\ref{fig:abund-case}). 
%\st{A slow warm-up time scale helps destroy ethanolamine by the surface reaction for extended periods.} 
It shows a peak abundance of 
%\st{$2.6 \times 10^{-10}$} 
$1.2 \times 10^{-9}$ around 90~K and then dramatically declines. 
%\st{Without reaction number R16, ethanolamine abundance drops to $\sim 7.4 \times 10^{-14}$ for hot corino case.}
It is worth noting that gas-phase ethanolamine production is predominantly influenced by ice-phase reaction R16 and its subsequent desorption.

We initially set the density at $3 \times 10^3$ cm$^{-3}$ and utilized low metal initial abundances, typically considered the standard for hot core/corino models. To assess the impact of the initial density on our calculation, we also experimented with an initial density of $10^4$ cm$^{-3}$. However, this alternative did not result in any noticeable deviation. This is because ethanolamine production is predominantly influenced during the warmup phase when the density is $10^7$ cm$^{-3}$.

\subsection{Astronomical implications}

Ethanolamine has been only detected so far towards a molecular cloud in the Galactic Center, G+0.693-0.027 \citep{Rivilla2021}, with an abundance compared to H$_2$ of $\sim 10^{-10}$. It has also been searched for toward star-forming regions, including several hot cores \citep{Widicus2003,Wirstrom2007}, and the prototypical hot corino IRAS 16293-2422 B \citep{naza24}, but the molecule was not detected, allowing only to compute upper limits for its abundance. 
\cite{Widicus2003} used the 10m telescope of the Caltech Submillimeter Observatory (CSO), and interferometric observations of the Owens Valley Radio Observatory (OVRO), while \cite{Wirstrom2007} used the Onsala 20m telescope. To convert the column density upper limits reported in these works into abundances we have used the values observed towards Orion KL. For the single-dish observations, we have corrected the beam dilution since it is expected that the spatial extent of the ethanolamine is compact as other complex molecules. 
For the size of the emission and the associated value of $N$(H$_2$) in the same volume, we have used the values derived for the Orion KL hot core by \cite{feng15} and \cite{croc14}, which are, respectively, 10\arcsec $\times$ 10\arcsec and $3.1 \times 10^{23}$~cm$^{-2}$, and 3.49\arcsec $\times$ 2.57\arcsec and $2.3 \times 10^{24}$~cm$^{-2}$. We obtain that the CSO observations provide an upper limit for the abundance of ethanolamine of $< (1.3-1.8) \times 10^{-9}$, while the Onsala observations give slightly higher values of $< (4.1-5.8) \times 10^{-9}$. 
However, we note that these numbers should be taken with caution, given the uncertainties of the size and $N$(H$_2$) assumptions. The OVRO observations of \cite{Widicus2003} were centered towards the Orion Compact Ridge, so in this case, we have used the size and $N$(H$_2$) derived by \cite{feng15} in this source, which is 3.49\arcsec $\times$ 2.57\arcsec and $1.28 \times 10^{24}$~cm$^{-2}$, respectively. The beam of the observations was 6.8\arcsec $\times$ 4.8\arcsec. After applying the beam dilution, we obtain an upper limit of ethanolamine abundance of $< 1.8 \times 10^{-10}$, which is at least one order of magnitude lower than that derived from the single-dish observations and very similar to the value calculated by \cite{Rivilla2021} towards G+0.693. 
Therefore, deep searches of ethanolamine towards hot cores are needed to establish if it is equally abundant than in G+0.693, or alternatively, this species has lower abundances in these environments. 
Our hot core model predicts a peak abundance of $\sim$ $2.8 \times 10^{-9}$.
Moreover, \cite{naza24} estimated an upper limit of ethanolamine column density 
$<6 \times 10^{14}$ cm$^{-2}$ towards IRAS 16293-2422 B using ALMA. 
Considering a H$_2$ column density of 
$\sim$ $10^{25}$~cm$^{-2}$ \citep{jorg16}, 
it yields an upper limit of abundance $< 6 \times 10^{-11}$. 
This indicates that the species is less abundant in this hot corino than in G+0.693. However, our hot corino model predicts a peak abundance of $\sim 1.2 \times 10^{-9}$.
For the sake of simplicity, 
%we estimated that the upper limit of the 
ethanolamine abundances in the surroundings of the hot core and hot corino region would vary between $10^{-11}-10^{-9}$.
This limit is highlighted with the yellow horizontal bar in Figure~\ref{fig:abund-case}.
%, indicating that our modeled abundance of ethanolamine aligns well with the reported upper limit values. 
Since the nondetection of ethanolamine towards hot cores/corinos was possibly due to a lack of sensitivity, more high sensitivity and high-resolution spectral data could lead to its future identification.
Furthermore, the high line density in these hot environments might be contributing to preventing the detection of ethanolamine, since its weak expected lines can be severely blended by stronger lines from more abundant species. To alleviate this, searches at lower frequencies, for instance in the new ALMA band 3, might help.

Apart from star-forming regions, ethanolamine has been detected
in the Almahata Sitta meteorite \citep{Glavin2010}. This suggests the presence of ethanolamine in solar system objects. The IR spectra presented for ethanolamine ices would help augment the JWST data toward understanding the origin of a phospholipid precursor in solar system objects.

\section{Conclusion} \label{sec:conclusion}

%This paper presents the temperature-dependent VUV and mid-IR spectra of ethanolamine ices under astrochemical conditions. We see a good agreement between the VUV and IR spectra vis-a-vis phase change (around 180~K) and sublimation (around 230~K). We observe that the phase change is strongly related to the deposition temperature and the isothermal condition. The VUV and IR spectral features presented in this paper can assist in the future unambiguous identification of ethanolamine in the ISM and other cold regions of the planetary and cometary bodies. Additionally, ethanolamine formation in high-mass and low-mass star-forming regions is explored by chemical models. It depicts that the formation of ethanolamine by RR surface reactions ($\rm{CH_2OH+NH_2CH_2\rightarrow NH_2CH_2CH_2OH}$ and $\rm{C_2H_4OH + NH_2\rightarrow NH_2CH_2CH_2OH}$) in warmer regions ($40-60$~K) is favorable. Our modeled abundances successfully explain the estimated upper limit of ethanolamine abundance in the surroundings of hot core/corino.

Theoretical and experimental work has been conducted to comprehend the formation of ethanolamine and its photochemistry under interstellar conditions. The major conclusions of this work are highlighted below:\\

\begin{itemize}

\item
In this study, we present the first temperature-dependent VUV and mid-IR spectra of pure ethanolamine ices under astrochemical conditions. Our results show a close correspondence between the VUV and IR spectra in relation to phase change (occurring around 180~K) and sublimation (occurring around 230~K). We have observed that the phase change is greatly influenced by the deposition temperature and the isothermal condition. The VUV and IR spectral features outlined in this paper can be valuable for its future identification of ethanolamine in the interstellar ices and other cold regions of planetary and cometary bodies.

\item Quantum chemical calculations are carried out to study the feasibility of ice-phase reactions involving the formation of ethanolamine. Some of these reactions contain activation barriers, which are estimated by our quantum chemical calculations. The obtained values are directly used in the model.

\item Ethanolamine formation in high-mass and low-mass star-forming regions is explored by chemical models. It depicts that the formation of ethanolamine by RR surface reactions in the warmer regions ($60-90$~K) is favorable. Our modeled abundances are consistent with the estimated upper limit of ethanolamine abundance in the hot core/corino surroundings.

\item 
To assist the interstellar search of several proposed precursors of ethanolamine (HNCCO, NH$_2$CHCHO, and NH$_2$CH$_2$CO), and to allow a proper derivation of the column densities from observations, we have performed theoretical calculations to derive their dipole moments, and their rotational constants. Upon reassessing the calculated rotational constants for ethanolamine and aligning them with prior data \citep{Penn1971}, we have extended our analysis to encompass the rotational and distortional constants for these species, which were absent in the literature.

\end{itemize}

\begin{acknowledgments}
\tiny{
% This work is completed while M.S. is a postdoctoral fellow at the Institute of Planetology and Astrophysics of Grenoble (IPAG), University of Grenoble Alpes.
We thank NSRRC for providing the beamtime and accessories that allowed us to perform the experiments. R.R., B.S., and N.J.M. are grateful to Sir John and Lady Mason Academic Trust for the support provided to carry out measurements at NSRRC. R.R., J.K.M, A.B., and B.S. acknowledge support from the Physical Research Laboratory (Department of Space, Government of India). R.R, J.K.M, and B.S would also like to acknowledge the Interdisciplinary program for astrobiology \& astrochemistry (IPAA) of Physical Research Laboratory. AB was a J.C. Bose Fellow during the period of this work. N.J. M. acknowledges the receipt of funding from the Europlanet 2024 RI, which has been funded by the European Union Horizon 2020 Research Innovation Programme under grant agreement No. 871149. M.S. acknowledges financial support through the European Research Council (consolidated grant COLLEXISM, grant agreement ID: 811363). A.D. acknowledges the support of MPE for sponsoring a scientific visit to MPE. V.M.R. acknowledges support from the grant PID2022-136814NB-I00 by the Spanish 
Ministry of Science, Innovation and Universities/State Agency of Research MICIU/AEI/10.13039/501100011033 and by ERDF, UE;  the grant RYC2020-029387-I funded by MICIU/AEI/10.13039/501100011033 and by "ESF, Investing in your future", and from the Consejo Superior de Investigaciones Cient{\'i}ficas (CSIC) and the Centro de Astrobiolog{\'i}a (CAB) through the project 20225AT015 (Proyectos intramurales especiales del CSIC); and from the grant CNS2023-144464 funded by MICIU/AEI/10.13039/501100011033 and by “European Union NextGenerationEU/PRTR”.}

\end{acknowledgments}

\bibliography{Ethanolamine.bbl}
\bibliographystyle{aasjournal}

%\section{Supplimentary material}

%%%%%%%%%%%%%%%%%%%%%%%%%%%%%%%%%%%%%%%%%%%%%%%%%%

%%%%%%%%%%%%%%%%% APPENDICES %%%%%%%%%%%%%%%%%%%%%

\appendix

\section{Gas phase destruction reactions of Ethanolamine and related species \label{sec:dest}}
The following reactions are included in our network for destroying gas phase ethanolamine and related species.
The format of this reaction follows the standard UMIST 2022 network. \\

\noindent :IN:H3+:HNCCO:NH3:HC2O+:::1:1.00e-09:-0.50:0.0:10:41000 \\
:IN:HCO+:HNCCO:HNCO:HC2O+:::1:1.00e-09:-0.50:0.0:10:41000 \\
:IN:He+:HNCCO:He:HC2O+:N::1:1.00e-09:-0.50:0.0:10:41000 \\
:IN:H+:HNCCO:NH:HC2O+:::1:1.00e-09:-0.50:0.0:10:41000 \\
:IN:C+:HNCCO:CN:HC2O+:::1:1.00e-09:-0.50:0.0:10:41000 \\
:IN:H3+:NH2CCO:NH3:HC2O+:H::1:1.00e-09:-0.50:0.0:10:41000 \\
:IN:HCO+:NH2CCO:HNCO:HC2O+:H::1:1.00e-09:-0.50:0.0:10:41000 \\
:IN:He+:NH2CCO:He:HC2O+:NH::1:1.00e-09:-0.50:0.0:10:41000 \\
:IN:H+:NH2CCO:NH2:HC2O+:::1:1.00e-09:-0.50:0.0:10:41000 \\
:IN:C+:NH2CCO:HCN:HC2O+:::1:1.00e-09:-0.50:0.0:10:41000 \\
:IN:H3+:NH2CHCO:NH3:HC2O+:H2::1:1.00e-09:-0.50:0.0:10:41000 \\
:IN:HCO+:NH2CHCO:HNCO:HC2O+:H2::1:1.00e-09:-0.50:0.0:10:41000 \\
:IN:He+:NH2CHCO:He:HC2O+:NH2::1:1.00e-09:-0.50:0.0:10:41000 \\
:IN:H+:NH2CHCO:NH3:HC2O+:::1:1.00e-09:-0.50:0.0:10:41000 \\
:IN:C+:NH2CHCO:H2CN:HC2O+:::1:1.00e-09:-0.50:0.0:10:41000 \\
:IN:H3+:NH2CH2CO:NH4+:CH:H2CO::1:1.00e-09:-0.50:0.0:10:41000 \\
:IN:HCO+:NH2CH2CO:NH3:HCO:HC2O+::1:1.00e-09:-0.50:0.0:10:41000 \\
:IN:He+:NH2CH2CO:NH3:He:HC2O+::1:1.00e-09:-0.50:0.0:10:41000 \\
:IN:H+:NH2CH2CO:NH3:HC2O+:H::1:1.00e-09:-0.50:0.0:10:41000 \\
:IN:C+:NH2CH2CO:NH3:HC2O+:C::1:1.00e-09:-0.50:0.0:10:41000 \\
:IN:H3+:NH2CH2CHO:NH4+:CH2:H2CO::1:1.00e-09:-0.50:0.0:10:41000 \\
:IN:HCO+:NH2CH2CHO:NH3:H2CO:HC2O+::1:1.00e-09:-0.50:0.0:10:41000 \\
:IN:He+:NH2CH2CHO:CH4+:He:HNCO::1:1.00e-09:-0.50:0.0:10:41000 \\
:IN:H+:NH2CH2CHO:NH3:HC2O+:H2::1:1.00e-09:-0.50:0.0:10:41000 \\
:IN:C+:NH2CH2CHO:NH3:HC2O+:CH::1:1.00e-09:-0.50:0.0:10:41000 \\
:IN:H3+:NH2C2H3OH:NH4+:CH3:H2CO::1:1.00e-09:-0.50:0.0:10:41000 \\
:IN:HCO+:NH2C2H3OH:NH2:CH3OH:HC2O+::1:1.00e-09:-0.50:0.0:10:41000 \\
:IN:He+:NH2C2H3OH:HCN:He:CH3OH2+::1:1.00e-09:-0.50:0.0:10:41000 \\
:IN:H+:NH2C2H3OH:NH4+:H2CO:CH::1:1.00e-09:-0.50:0.0:10:41000 \\
:IN:C+:NH2C2H3OH:NH4+:H2CO:C2::1:1.00e-09:-0.50:0.0:10:41000 \\
:IN:H3+:NH2CH2CH2O:NH4+:CH3:H2CO::1:1.00e-09:-0.50:0.0:10:41000 \\
:IN:HCO+:NH2CH2CH2O:NH2+:CH3CHO:HCO::1:1.00e-09:-0.50:0.0:10:41000 \\
:IN:He+:NH2CH2CH2O:NH2+:He:CH3CHO::1:1.00e-09:-0.50:0.0:10:41000 \\
:IN:H+:NH2CH2CH2O:NH4+:H2CO:CH::1:1.00e-09:-0.50:0.0:10:41000 \\
:IN:C+:NH2CH2CH2O:NH4+:H2CO:C2::1:1.00e-09:-0.50:0.0:10:41000 \\
:IN:H3+:NH2CH2CH2OH:NH3:CH5+:H2CO::1:1.00e-09:-0.50:0.0:10:41000 \\
:IN:HCO+:NH2CH2CH2OH:OCN:CH3:CH3OH2+::1:1.00e-09:-0.50:0.0:10:41000 \\
:IN:He+:NH2CH2CH2OH:H2CN:He:CH3OH2+::1:1.00e-09:-0.50:0.0:10:41000 \\
:IN:H+:NH2CH2CH2OH:NH4+:H2CO:CH2::1:1.00e-09:-0.50:0.0:10:41000 \\
:IN:C+:NH2CH2CH2OH:NH4+:H2CO:C2H::1:1.00e-09:-0.50:0.0:10:41000 \\
:IN:H3+:NH2CH:NH4+:CH2:::1:1.00e-09:-0.50:0.0:10:41000:E:C:: \\
:IN:HCO+:NH2CH:NH4+:C2O:::1:1.00e-09:-0.50:0.0:10:41000:E:C:: \\
:IN:He+:NH2CH:NH3:C+:He::1:1.00e-09:-0.50:0.0:10:41000:E:C:: \\
:IN:H+:NH2CH:NH4+:C:::1:1.00e-09:-0.50:0.0:10:41000:E:C:: \\
:IN:C+:NH2CH:NH3:C2+:::1:1.00e-09:-0.50:0.0:10:41000:E:C:: \\
:IN:H3+:NH2CH2:NH4+:CH3:::1:1.00e-09:-0.50:0.0:10:41000:E:C:: \\
:IN:HCO+:NH2CH2:NH4+:C2O:H::1:1.00e-09:-0.50:0.0:10:41000:E:C:: \\
:IN:He+:NH2CH2:NH3:CH+:He::1:1.00e-09:-0.50:0.0:10:41000:E:C:: \\
:IN:H+:NH2CH2:NH4+:CH:::1:1.00e-09:-0.50:0.0:10:41000:E:C:: \\
:IN:C+:NH2CH2:NH3:C2H+:::1:1.00e-09:-0.50:0.0:10:41000:E:C:: \\
:IN:H3+:C3H7NO2:NH4+:CH2CO:CH3OH::1:1.00e-09:-0.50:0.0:10:41000 \\
:IN:H+:C3H7NO2:NH4+:CH3CHO:CO::1:1.00e-09:-0.50:0.0:10:41000 \\
:IN:H3+:C2H4OH:H3O+:C2H5:::1:1.00e-09:-0.50:0.0:10:41000:E:C:: \\
:IN:H+:C2H4OH:H3O+:C2H3:::1:1.00e-09:-0.50:0.0:10:41000:E:C:: \\
:IN:HCO+:C2H4OH:H3O+:C2H3:CO::1:1.00e-09:-0.50:0.0:10:41000:E:C:: \\
:IN:He+:C2H4OH:He:C2H4:OH+::1:1.00e-09:-0.50:0.0:10:41000:E:C:: \\
:IN:C+:C2H4OH:HCO+:C2H4:::1:1.00e-09:-0.50:0.0:10:41000:E:C:: \\
:CR:HNCCO:CRPHOT:NH:C2O:::1:1.30e-17:0.00:1500.0:10:41000:M:C \\
:CR:NH2CCO:CRPHOT:NH2:C2O:::1:1.30e-17:0.00:1500.0:10:41000:M:C \\
:CR:NH2CHCO:CRPHOT:NH3:C2O:::1:1.30e-17:0.00:1500.0:10:41000:M:C \\
:CR:NH2CH2CO:CRPHOT:NH2:CH2CO:::1:1.30e-17:0.00:1500.0:10:41000:M:C \\
:CR:NH2CH2CHO:CRPHOT:NH2:CH2CHO:::1:1.30e-17:0.00:1500.0:10:41000:M:C \\
:CR:NH2CH2CH2O:CRPHOT:NH2:CH3CHO:::1:1.30e-17:0.00:1500.0:10:41000:M:C \\
:CR:NH2CH2CH2OH:CRPHOT:NH3:CH3CHO:::1:1.30e-17:0.00:1500.0:10:41000:M:C \\
:CR:NH2C2H3OH:CRPHOT:NH2:CH3CHO:::1:1.30e-17:0.00:1500.0:10:41000:M:C \\
:CR:NH2CH:CRPHOT:NH2:CH:::1:1.30e-17:0.00:1500.0:10:41000:M:C \\
:CR:NH2CH2:CRPHOT:NH2:CH2:::1:1.30e-17:0.00:1500.0:10:41000:M:C \\
:CR:C3H7NO2:CRPHOT:H2:CH3CHO:HNCO::1:1.30e-17:0.00:1500.0:10:41000:M:C \\
:CR:C2H4OH:CRPHOT:C2H4:OH:::1:1.30e-17:0.00:1500.0:10:41000:M:C \\
:PH:HNCCO:PHOTON:NH:C2O:::1:1.00e-10:0.00:2.0:10:41000:M:C \\
:PH:NH2CCO:PHOTON:NH2:C2O:::1:1.00e-10:0.00:2.0:10:41000:M:C \\
:PH:NH2CHCO:PHOTON:NH3:C2O:::1:1.00e-10:0.00:2.0:10:41000:M:C \\
:PH:NH2CH2CO:PHOTON:NH2:CH2CO:::1:1.00e-10:0.00:2.0:10:41000:M:C \\
:PH:NH2CH2CHO:PHOTON:NH2:CH2CHO:::1:1.00e-10:0.00:2.0:10:41000:M:C \\
:PH:NH2CH2CH2O:PHOTON:NH2:CH3CHO:::1:1.00e-10:0.00:2.0:10:41000:M:C \\
:PH:NH2CH2CH2OH:PHOTON:NH3:CH3CHO:::1:1.00e-10:0.00:2.0:10:41000:M:C \\
:PH:NH2C2H3OH:PHOTON:NH2:CH3CHO:::1:1.00e-10:0.00:2.0:10:41000:M:C \\
:PH:NH2CH:PHOTON:NH2:CH:::1:1.00e-10:0.00:2.0:10:41000:M:C \\
:PH:NH2CH2:PHOTON:NH2:CH2:::1:1.00e-10:0.00:2.0:10:41000:M:C \\
:PH:C3H7NO2:PHOTON:H2:CH3CHO:HNCO::1:1.00e-10:0.00:2.0:10:41000:M:C \\
:PH:C2H4OH:PHOTON:C2H4:OH:::1:1.00e-10:0.00:2.0:10:41000:M:C

\end{document}